\documentclass[11pt,letterpaper]{article}
\usepackage[T1]{fontenc}
\usepackage[margin=1in]{geometry}
\usepackage{amsmath,amssymb,amsfonts,amsthm,bbold}
\usepackage{graphicx,array,multirow,rotating}
\usepackage[section]{placeins}
\usepackage{enumitem}
\usepackage{tikz}
\usetikzlibrary{shapes.geometric,positioning}
\usepackage[lined,boxed,ruled,linesnumbered]{algorithm2e}
\usepackage[authoryear,round]{natbib}
\usepackage[hidelinks,hypertexnames=false]{hyperref}
\hypersetup{pdftitle={Rematching Estimators for Average Treatment Effects},
 pdfauthor={Lam Lam Hui and Kin Wai Chan}}
\graphicspath{{art/}{art_supp/}}
\allowdisplaybreaks
\newcommand{\bc}{\textnormal{bc}}
\newcommand{\re}{\textnormal{re}}
\newcommand{\pr}{\mathrm{pr}}
\newcommand{\E}{E}
\newcommand{\cov}{\mathrm{cov}}
\newcommand{\var}{\mathrm{var}}
\newcommand{\T}{\textsc{T}}
\newcommand{\MSE}{\mathop{\mathrm{MSE}}}
\newcommand{\Bern}{\textnormal{Bern}}
\newcommand{\Normal}{\textnormal{N}}
\newcommand{\Bin}{\textnormal{Bin}}
\newcommand{\Beta}{\textnormal{Beta}}
\newcommand{\Unif}{\textnormal{Unif}}
\DeclareMathOperator*{\argmin}{arg\,min}
\DeclareMathOperator*{\argmax}{arg\,max}
\newtheorem{theorem}{Theorem}
\newtheorem{lemma}{Lemma}

\newtheorem{proposition}{Proposition}
\theoremstyle{definition}

\newtheorem{remark}{Remark}
\newtheorem{assumption}{Assumption}
\numberwithin{equation}{section}

\title{Rematching Estimators for Average Treatment Effects}
\author{Lam Lam Hui and Kin Wai Chan\\[4pt]
\small Department of Statistics and Data Science\\
\small The Chinese University of Hong Kong, Shatin, Hong Kong\\[4pt]
\small \texttt{lamlamhui@link.cuhk.edu.hk} and \texttt{kinwaichan@cuhk.edu.hk}}
\date{}
\begin{document}
\maketitle
\begin{abstract}
Matching estimators are widely applied in practice for their great intuitive appeal.
However, simple matching estimators with a fixed number of matches ($M_0$) are generally inefficient.
In this article, we propose matching estimators with a variable number of matches to gain efficiency via rematching.
Rather than increasing $M_0$ to gain precision,
which introduces an increase in bias, the key is to rematch the treated units
from the opposite direction to utilize unmatched control units.
Our rematching estimators are applicable to both the average treatment effect
and its counterpart for the treated population.
The proposed rematching estimators are proven asymptotically valid and
uniformly more efficient than matching estimators with the same $M_0$.
Simulation results confirm that the proposed rematching estimators substantially improve
the simple matching estimators in finite samples.
As an empirical illustration, we apply the estimators proposed
in this article to the National Supported Work data.
\end{abstract}

\noindent\textbf{Keywords:} Average treatment effects, causal inference, potential outcome.

\section{Introduction}
Matching estimators are widely used to estimate the average effects of a program, medical treatment, or policy intervention by empirical researchers because of their great intuitive appeal.
In contrast to other studies on propensity score matching estimators \citep{rosenbaum1983central}, the term `matching estimator' in this article is reserved for estimators that match each unit to a number of units in the opposite treatment group by multivariate distance matching.
Under the setting of matching with replacement,
\cite{abadie2006large,abadie2008failure,abadie2011bias,abadie2012martingale} conducted various research on simple matching estimators with a fixed number of matches.
While simple matching estimators are commonly used in practice,
research has shown that some undesirable properties exist for simple matching estimators \citep{abadie2006large}.
In contrast to some regression adjustment estimators (e.g., \cite{hahn1998role}; \cite{heckman1998matching}; \cite{imbens2005mean};
\cite{chen2008semiparametric}) and weighting estimators (e.g., \cite{horvitz1952generalization}; \cite{robins1995semiparametric}; \cite{hirano2003efficient}; \cite{abadie2005semiparametric}), matching estimators may not be fully efficient \citep{abadie2006large}.
According to a recent work by \cite{lin2023estimation},
when matching is performed on a vector of covariates,
allowing $M_0\rightarrow\infty$ with the sample size leads to the semiparametric efficiency
of bias-corrected matching estimators if the outcome model is appropriately specified.
However, when the outcome model is misspecified,
increasing the number of matched units through matching in the
original direction may introduce matches of lower quality leading to a bias increase,
which may render the causal interpretation of the estimates invalid.
It motivates us to develop a new matching procedure that better utilizes observations and gains efficiency without increasing the bias too much.

The major contribution of this article is that we can achieve variance reduction by incorporating an additional rematching step with a small value of $M_0$,
as opposed to using a much larger $M_0$ in a standard matching estimator.
The use of a smaller $M_0$ ensures high-quality matching units, leading to a smaller bias.
Based on a landmark paper by \cite{abadie2006large}, which provided the first large sample analysis of simple matching estimators, we show the asymptotic validity of our rematching-based estimators when matched on a scalar covariate.
By incorporating a bias correction method by \cite{abadie2011bias}, our method can be asymptotically valid even for matching based on a vector of covariates.
As the simulation results show, our bias-corrected rematching estimators lead to a substantial reduction in variance compared with bias-corrected simple matching estimators
when the bias is negligible after correction and have a lower bias when the outcome model is misspecified.
In summary, our proposed rematching estimators with a small $M_0$ are better alternatives to the matching estimators with a large $M_0$.

\section{Background and setup}\label{sec:setup}
Matching estimators are often used to estimate treatment effects in observational studies where experimental data are unavailable.
And the now dominant approach to analysing causal effects in observational studies was formulated by \cite{rubin1973matching, rubin1973use, rubin1974estimating, rubin1977assignment, rubin1978bayesian}.
For $N$
observed units, indexed by $i = 1,\ldots, N$,
the $i$th outcome variable is
\[Y_i =
\begin{cases}
Y_i(0)  & \text{if } D_i = 0; \\
Y_i(1) & \text{if } D_i = 1, \\
\end{cases}
\] where $D_i$ indicates the treatment received ($D_i = 1$ if treated and $D_i = 0$ otherwise), and $Y_i(d)$ is the potential outcome under $D_i=d$.
The sizes of the control and the treated groups are $N_0$ and $N_1$, respectively, with $N_0 + N_1 = N$.
A vector of covariates $X_i$ is also observed for the $i$th unit.
Based on $\{(Y_i, D_i, X_i)\}^N_{i=1}$,  we conduct the inference on the average treatment effect \[\tau = \E\{Y_i(1) - Y_i(0)\},\] and its counterpart for the treated population
defined as
\[\tau^t = \E\{Y_i(1) - Y_i(0)\mid D_i=1\}.\]

In this article, we consider continuous covariates only and ignore the possibility of ties.
Let $\mathbb{1}\{\cdot\}$ be the indicator function and $\lVert \cdot \rVert$ be the Euclidean norm.
By simple matching estimators, we estimate the potential outcomes by
\begin{align*}
\hat{Y}_i(0) =
\begin{cases}
Y_i & \text{if } D_i = 0;\\
M_0^{-1}\sum_{j\in \mathcal{J}_{M_0}(i)} Y_j & \text{if } D_i = 1,
\end{cases}\\
\quad
\hat{Y}_i(1) =
\begin{cases}
M_0^{-1}\sum_{j\in \mathcal{J}_{M_0}(i)} Y_j & \text{if } D_i = 0;\\
Y_i & \text{if } D_i = 1,\\
\end{cases}
\end{align*}
where
\begin{equation}
\mathcal{J}_{M_0}(i) = \{j_1(i),\ldots, j_{M_0}(i)\}
\label{eqt:JMF}
\end{equation}
is the set of indices corresponding to the closest $M_0$ matches for the unit $i$ in the opposite treatment group, i.e.,
$j_m(i)$ is the index $j \in  \{1,\ldots, N\}$ that solves $D_j = 1 - D_i$ and
\[
    \sum_{\ell:D_\ell=1-D_i} \mathbb{1}\{\lVert X_\ell- X_i\rVert  \leq
    \lVert X_j - X_i \rVert \} = m \quad (m=1,\ldots,M_0).
\]
Let  $K_{M_0}(i)$ denotes the number of times unit $i$ is matched, i.e.,
\begin{equation}
    K_{M_0}(i) = \sum^N_{\ell=1}\mathbb{1}\{i \in \mathcal{J}_{M_0}(\ell)\}.
    \label{eqt:KM0}
\end{equation}
The average treatment effects are estimated by
\begin{align}
\hat{\tau}_{0}
    &= \frac{1}{N}\sum^N_{i=1}{\hat{Y}_i(1) -\hat{Y}_i(0)}
    =\frac{1}{N}\sum_{i=1}^N(2D_i-1)\left\{1+\frac{K_{M_0}(i)}{M_0}\right\}Y_i\label{eqt:hTauF}, \\
\hat{\tau}_{0}^{t}
    &= \frac{1}{N_1}\sum_{\substack{1\le i\le N \\ D_i=1}}{Y_i -\hat{Y}_i(0)}
    = \frac{1}{N_1}\sum_{i=1}^N \left\{D_i - (1-D_i)
    \frac{K_{M_0}(i)}{M_0}\right\}Y_i.\nonumber
\end{align}
To remove the bias of matching, \cite{abadie2011bias} combined matching with a bias correction proposed in \cite{rubin1973use} and \cite{quade1982nonparametric}, which produces bias-corrected matching estimators $\hat{\tau}_{0,\bc}$ and  $\hat{\tau}^{t}_{0,\bc}$.

\begin{figure}[t!]
\begin{center}
\newcommand{\markerr}{1.1}
\newcommand{\triAlpha}{2}
  \begin{tikzpicture}[thick,scale=0.9, every node/.style={scale=0.9},
blacknode/.style={circle, draw=black!70, fill=black!70, ultra thin, minimum size=1mm},
rednode/.style={
    minimum size=4 mm, inner sep=0pt, draw=none, fill=none,
    path picture={
      \begin{scope}[x=1mm,y=1mm, rotate=90]
        \pgfmathsetmacro{\R}{\markerr*\triAlpha}
        \path ( 90:\R) coordinate (A)
              (-30:\R) coordinate (B)
              (210:\R) coordinate (C);
        \fill[red!70] (A) -- (B) -- (C) -- cycle;
      \end{scope}
    }
  },
]
\draw[dashed, thick](0.67,-.9) ellipse (1.8 and 1.25);
\draw[dashed, thick](1.56,-.5) ellipse (1.38 and 1.85);
\draw[rotate=22][dashed, thick](3.35,-1.7) ellipse (1.2 and 1.8);

\node[blacknode](t1){} node[above left=-1mm of t1]{$1$};
\node[blacknode](t3)[above=5mm of t1, xshift=4cm]{} node[above right=-1.5mm of t3]{$3$};
\node[blacknode](t2)[right=7mm of t1, yshift=1cm]{} node[above left=-1mm of t2]{$2$};

\node[rednode](c1)[right=2mm of t3, yshift=-2.5cm]{} node[below right=-2.5mm of c1]{$12$};
\node[rednode](c2)[left=3mm of c1, yshift=3mm]{} node[below=-1mm of c2]{$11$};
\node[rednode](c3)[right=2mm of c2, yshift=5mm]{} node[above=-1mm of c3]{$13$};

\node[rednode](c4)[left=3mm of c2, yshift=.8cm]{} node[right=-1.5mm of c4]{$10$};
\node[rednode](c5)[left=0mm of c4, yshift=5mm]{} node[right=-1mm of c5]{$9$};
\node[rednode](c6)[left=5mm of c4, yshift=-2mm]{} node[above left=-3mm of c6]{$8$};

\node[rednode](c7)[right=1mm of t2, yshift=-2.1cm]{} node[below=-1mm of c7]{$7$};
\node[rednode](c8)[left=1mm of t2, yshift=-2.5cm]{} node[right=-1mm of c8]{$6$};
\node[rednode](c9)[left=2mm of t2, yshift=-2cm]{} node[above=-1mm of c9]{$5$};

\node[rednode](c10)[below= of t1, xshift=-.6cm]{} node[above=-1mm of c10]{$4$};

\draw[->, thick] (-1.2,-2.5)--(5.2,-2.5) node[right]{$X$};
\draw[->, thick] (-1.2,-2.5)--(-1.2,1.8) node[above]{$Y$};

\end{tikzpicture}

\caption{An example of the simple matching procedure that finds matches from the controls (triangles) for each treated unit (circles) with the horizontal axis and the vertical axis representing the scalar covariate $X$ and the outcome $Y$, respectively.
The markers represent the values of $(X_i,Y_i)$, $i=1,\ldots,13$.
We set $M_0=5$ so that each control is matched at least once.
When $\tau^t$ is the estimand of interest, the matching results are represented by dotted loops.
}
\label{fig:lowQualityMatches}
\end{center}
\end{figure}
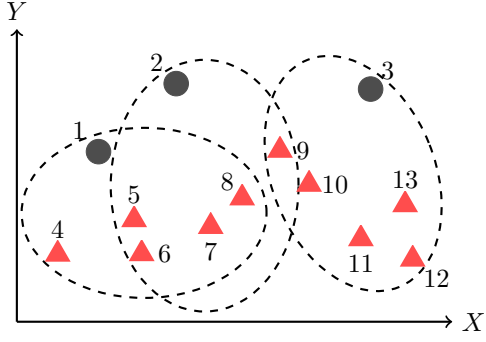

Despite being commonly used by practitioners,
the matching estimators $\hat{\tau}_{0}$ and $\hat{\tau}_{0}^t$ are not free of problems.
It is found that when there is a larger reservoir of potential controls than treated and $M_0$ is not large enough, the matching estimators are inefficient since some control units will likely be discarded, leading to an efficiency loss.
However, increasing $M_0$ for precision may repeatedly include far away units as low-quality matches, accompanied by a significant bias increase if the bias is not well-corrected; see Figure \ref{fig:lowQualityMatches}.
It thus motivates us to revisit the matching procedure
and propose a new generation of matching estimators to gain efficiency without increasing the bias too much through better utilization of unmatched control units.

\section{Rematching estimators}\label{sec:proposal}
\subsection{A matching-and-rematching procedure}

We propose rematching estimators to avoid efficiency loss due to points discarding when there is a large reservoir of potential controls.
The efficiency gain is obtained without increasing the bias too much by rematching the treated units from an opposite direction.
The discussion in this section is based on $\tau^t$.
Because often, matching estimators have been used when (1) the interest is in $\tau^t$, and (2) a large reservoir of potential controls is available \citep{imbens2009recent}.

\begin{figure}[t!]
\begin{center}
\newcommand{\markerr}{1.1}
\newcommand{\triAlpha}{2}
  \begin{tikzpicture}[thick,scale=0.85, every node/.style={scale=0.85},
blacknode/.style={circle, draw=black!70, fill=black!70, ultra thin, minimum size=1mm},
rednode/.style={
    minimum size=4 mm, inner sep=0pt, draw=none, fill=none,
    path picture={
      \begin{scope}[x=1mm,y=1mm, rotate=90]
        \pgfmathsetmacro{\R}{\markerr*\triAlpha}
        \path ( 90:\R) coordinate (A)
              (-30:\R) coordinate (B)
              (210:\R) coordinate (C);
        \fill[red!70] (A) -- (B) -- (C) -- cycle;
      \end{scope}
    }
  },
]

\draw[rotate=72][dashed, thick](-.8,-.4) ellipse (1.4 and 1);
\draw[dashed, thick](1.1,-.4) ellipse (.9 and 1.82);
\draw[rotate=5][dashed, thick](4.2,-.7) ellipse (.8 and 1.8);

\draw[rotate=72][dashed, thick](1.6,-7.69) ellipse (1.4 and 1);
\draw[dashed, thick](8.77,-.4) ellipse (.9 and 1.82);
\draw[rotate=5][dashed, thick](11.85,-1.4) ellipse (.8 and 1.8);

\draw[rotate=33][thick](7.95,-5) ellipse (.42 and 1.5);
\draw[rotate=0][thick](11.1,.35) ellipse (1.1 and 1.1);

\node[blacknode](t1){} node[above left=0mm of t1]{$1$};
\node[blacknode](t3)[above=5mm of t1, xshift=4cm]{} node[above left=0mm of t3]{$3$};
\node[blacknode](t2)[right=7mm of t1, yshift=1cm]{} node[above left=1mm of t2]{$2$};

\node[rednode](c1)[right=2mm of t3, yshift=-2.5cm]{} node[below right=-2mm of c1]{$12$};
\node[rednode](c2)[left=3mm of c1, yshift=3mm]{} node[below left=-2mm of c2]{$11$};
\node[rednode](c3)[right=2mm of c2, yshift=5mm]{} node[above=0mm of c3]{$13$};

\node[rednode](c4)[left=3mm of c2, yshift=.8cm]{} node[below=-1mm of c4]{$10$};
\node[rednode](c5)[left=0mm of c4, yshift=8mm]{} node[above=0mm of c5]{$9$};
\node[rednode](c6)[left=3mm of c4, yshift=-2mm]{} node[above = -0.5mm of c6]{$8$};

\node[rednode](c7)[right=0mm of t2, yshift=-2.1cm]{} node[above=0mm of c7]{$7$};
\node[rednode](c8)[left=1mm of t2, yshift=-2.5cm]{} node[below right =-2mm of c8]{$6$};
\node[rednode](c9)[left=2mm of t2, yshift=-2cm]{} node[above=-0.5mm of c9]{$5$};

\node[rednode](c10)[below= of t1, xshift=-.6cm]{} node[above=-0.5mm of c10]{$4$};

\node[blacknode](t1p)[right=6.2cm of t1]{};
\node[blacknode](t3p)[above=5mm of t1p, xshift=4cm]{};
\node[blacknode](t2p)[right=7mm of t1p, yshift=1cm]{};

\node[rednode](c1p)[right=2mm of t3p, yshift=-2.5cm]{};
\node[rednode](c2p)[left=3mm of c1p, yshift=3mm]{};
\node[rednode](c3p)[right=2mm of c2p, yshift=5mm]{};

\node[rednode](c4p)[left=3mm of c2p, yshift=.8cm]{};
\node[rednode](c5p)[left=0mm of c4p, yshift=8mm]{};
\node[rednode](c6p)[left=3mm of c4p, yshift=-2mm]{};

\node[rednode](c7p)[right=0mm of t2p, yshift=-2.1cm]{};
\node[rednode](c8p)[left=1mm of t2p, yshift=-2.5cm]{};
\node[rednode](c9p)[left=2mm of t2p, yshift=-2cm]{};

\node[rednode](c10)[below= of t1p, xshift=-.6cm]{};

\draw[->, thick] (-1.2,-2.5)--(5.2,-2.5) node[right]{$X$};
\draw[->, thick] (-1.2,-2.5)--(-1.2,1.8) node[above]{$Y$};

\draw[->, thick] (6.5,-2.5)--(12.9,-2.5) node[right]{$X$};
\draw[->, thick] (6.5,-2.5)--(6.5,1.8) node[above]{$Y$};

\draw[black,  thick, ->] (5.5,-0.35) -- (6.2,-0.35);

\end{tikzpicture}
\caption{An example of the matching-and-rematching procedure that finds matches from the controls (triangles) for each treated unit (circles) with the horizontal axis and the vertical axis representing the scalar covariate $X$ and the outcome $Y$, respectively.
The markers represent the values of $(X_i,Y_i)$, $i=1,\ldots,13$.
When $M_0=3$ and $\tau^t$ is the estimand, the results of the first and the second matching are represented by dotted and solid loops, respectively.
}
\label{fig:eg}
\end{center}
\end{figure}
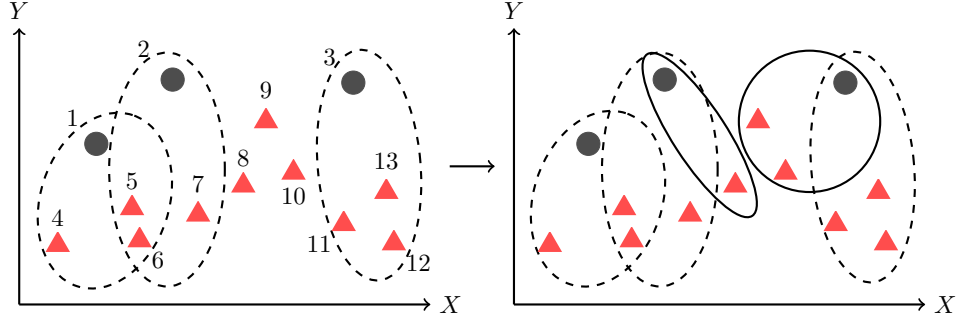

Figure \ref{fig:eg} is a visualization of the whole matching process that produces our proposed estimators.
After the first matching, which is the simple matching,
a dotted loop will cover both a control unit $i$ and a treated unit $\ell$ if the unit $i$ is matched to unit $\ell$ in the matching,
e.g., the 1st treated unit is matched to the 4th, 5th, and 6th controls if $M_0=3$.
However, no dotted loops will cover unit $i$ if it remains unmatched after the first matching, e.g., the 8th, 9th, and 10th units.
When there is a large reservoir of potential controls, some controls will likely be discarded without being used, leading to an efficiency loss.
To gain efficiency, we propose to rematch the treated units after the first matching by matching those unmatched controls to them, which means we do the rematching from an opposite direction.
Instead of performing $M_0$-nearest neighbour search for each fixed treated unit as in the first matching process,
we find only the nearest neighbour for each fixed unmatched control.
Because though matching for unmatched points can provide efficiency gain, those unmatched points are generally not matches of good quality and should not be overused.
The final matching results after rematching are in the right panel of Figure \ref{fig:eg}.
As an example of the rematching results represented by solid loops, the 9th and 10th units are rematched to the 3rd unit which is the closest to the 9th and the 10th units.

The rematching step provides a channel to extract additional information
from the observations that are originally unused.
Intuitively, it has two appealing features.
First, the rematching step utilizes more observations that
may lead to a reduction in variance.
Second, the rematching step
tends to match higher-quality observations in the opposite direction than lower-quality matches in the original matching direction.
Consequently, by construction, the resulting estimator has the potential to strike a good balance between variance and bias.

Regarding the choice of $M_0$, ``little is known about the optimal number of matches, or about data-dependent ways of choosing it"  \citep{imbens2009recent}.
It is suggested that people fix $M_0$ at one when using rematching estimators on finite samples as empirical researchers typically do when using matching estimators.
In the asymptotic context, increasing $M_0$ with the sample size, which shares a similar idea to that of \cite{lin2023estimation} is advisable for rematching estimators.
However, we do not suggest a diverging $M_0$ considering the large bias when the outcome model is not well-specified.
More attention is attached to the particular choice of $M_0=1$ in this article as our simulated data and real data are of small sample sizes.

\subsection{Main proposal}
Let $\emptyset$ be an empty set. Every unmatched unit is matched to its nearest neighbour with the index
\begin{eqnarray}\label{eqt:JU}
\mathcal{J}_{\re}(i)=
\begin{cases}
\mathcal{J}_1(i) & \text{if } K_{M_0}(i) = 0;\\
\emptyset & \text{if } K_{M_0}(i) \neq 0.\\
\end{cases}
\end{eqnarray}
Let $M_{\re}(\ell)$ be the number of times unit $\ell$ is matched to unmatched units, i.e.,
\begin{equation}
    M_{\re}(\ell) = \sum^N_{i=1}\mathbb{1}\{\ell \in \mathcal{J}_{\re}(i)\}.
    \label{eqt:KU}
\end{equation}
The rematching process increases the number of matched samples in the control group by matching each treated unit $\ell$ to a set of controls of a variable size  \[M(\ell) = M_0 + M_{\re}(\ell),\] where $M_0$ is the fixed number of nearest neighbours found in the first matching and $M_{\re}(\ell)$ is the variable number of unmatched units matched to unit $\ell$ in the rematching.
Then, an equal matching weight of $1/M(\ell)$ will be given to all control units that are matched to unit $\ell$, which deviates the matching weight of those $M(\ell)$ controls from $1/M_0$ or 0 to a non-zero variable value; see Remark \ref{rem:unequalWeight}.
We now define the weighted version of the total number of times the unit $i$ is matched to treated units as
\begin{align}
    K(i) = \left[ \sum^N_{\ell=1}\mathbb{1}\{i \in \mathcal{J}_{M_0}(\ell)\}  \frac{M_0}{M_{\re}(\ell)+M_0} \right]& \nonumber\\
+\mathbb{1}\{K_{M_0}(i)=0\}&  \frac{M_0}{M_{\re}(\mathcal{J}_{\re}(i))+M_0}
\label{eqt:KM}
\end{align}
so that $K(i)/M_0$ is the total matching weight assigned to unit $i$.
If there is $\ell\in [1,N]$ such that $ \mathbb{1}\{i \in \mathcal{J}_{M_0}(\ell)\}=1$, or equivalently, $K_{M_0}(i)\neq0$,
then unit $i$ is matched in the first matching, and after rematching, $K(i)/M_0=\sum^N_{\ell=1}\mathbb{1}\{i \in \mathcal{J}_{M_0}(\ell)\}/\{M_{\re}(\ell)+M_0\}$.
Otherwise, the originally unmatched unit $i$ will have a matching weight of $1/\{M_{\re}(\mathcal{J}_{\re}(i))+M_0\}$ as it is rematched to unit $\mathcal{J}_{\re}(i)$.
Define the set of indices of unmatched units that unit $i$ is rematched to as
\[
\mathcal{L}_{\re}(i) = \{\ell \in  \{1,\ldots, N\} : i \in \mathcal{J}_{\re}(\ell)
\}.
\]
The set of indices of units that unit $i$ is matched to is then given by
\[\mathcal{J}(i)=\mathcal{J}_{M_0}(i)\cup\mathcal{L}_{\re}(i).\]
The following proposition gives an equivalent form of $K(i)$, which is concise and accessible.
\begin{proposition}\label{prop:1}
The term $K(i)$ $(i = 1,\ldots,N)$ in (\ref{eqt:KM}) can be equivalently computed as
\[
    K(i) = \sum^N_{\ell=1}\mathbb{1}\{i \in \mathcal{J}(\ell)\}  \frac{M_0}{M(\ell)}.
\]
\end{proposition}
Under the proposed matching-and-rematching scheme, the potential outcomes are estimated as
\begin{align*}
\Tilde{Y}_{i}(0) &=
\begin{cases}
Y_i & \text{if } D_i = 0;\\
M(i)^{-1}\sum_{j\in \mathcal{J}(i)} Y_j & \text{if } D_i = 1,\\
\end{cases}
\\
\Tilde{Y}_{i}(1) &=
\begin{cases}
M(i)^{-1}\sum_{j\in \mathcal{J}(i)} Y_j & \text{if } D_i = 0;\\
Y_i & \text{if } D_i = 1.\\
\end{cases}
\end{align*}
\begin{remark}\label{rem:unequalWeight}
Besides giving an equal weight of $1/M(\ell)$ to controls matched to unit $\ell$,
there are some possible methods to treat the $M_0$ matches and the $M_{\re}(\ell)$ matches differently,
which prevents the rematching process from incorporating large biases produced by matches of poor quality.
For instance, we may use a kernel function
on the $M_{\re}(\ell)$ matches (or all matches) to give larger weights to controls with smaller distances.
Since it is beyond the scope of this article, we leave this technical modification for future study.

\end{remark}
Our proposed rematching estimators for average treatment effects $\tau$ and $\tau^t$ are given by
\[
    \hat{\tau} =
    \frac{1}{N}\sum^N_{i=1}{\Tilde{Y}_i(1) -\Tilde{Y}_i(0)}, \quad
    \hat{\tau}^t = \frac{1}{N_1}\sum_{\substack{1\le i\le N \\ D_i=1}}{Y_i -\Tilde{Y}_i(0)},
\]
respectively.
Both of which average within-match differences in the potential outcome between the treated and the control units.
Proposition \ref{prop:2} below provides equivalent forms of $\hat{\tau}$ and $\hat{\tau}^t$.
\begin{proposition}\label{prop:2}
    The rematching estimators can be equivalently written as
\begin{align}
    \hat{\tau}
    &= \frac{1}{N}\sum_{i=1}^N(2D_i-1)\left\{1+\frac{K(i)}{M_0}\right\}Y_i, \label{eqt:tauHat}
    \\
    \hat{\tau}^t
    &=  \frac{1}{N_1}\sum_{i=1}^N \left\{D_i - (1-D_i)
     \frac{K(i)}{M_0}\right\}Y_i.
     \label{eqt:tauTHat}
\end{align}
\end{proposition}
Denote $  \mu(d, x) = \E(Y\mid D = d, X = x).$
When the bias correction is needed, we follow the study of \cite{abadie2011bias} and estimate the bias as follows:
\begin{align*}
\hat{B} &= \frac{1}{N} \sum_{i=1}^N
    \frac{(2D_i - 1)}{M(i)}\sum_{j\in \mathcal{J}_M (i)}\{ \hat{\mu}(1 - D_i, X_i) - \hat{\mu}(1 - D_i, X_j)\},\\
    \hat{B}^t &= \frac{1}{N_1} \sum_{i=1}^N
    \frac{D_i}{M(i)}\sum_{j\in \mathcal{J}_M (i)}\{ \hat{\mu}(0, X_i) - \hat{\mu}(0, X_j)\},
\end{align*}
where $\hat{\mu}(d, x)$ $(d=0,1)$ is a non-parametric series regression estimator estimating $\mu(d, x)$.
Then the bias-corrected rematching estimators are
\[
    \hat{\tau}_{\bc}=\hat{\tau}-\hat{B},
    \quad
    \hat{\tau}^t_{\bc}=\hat{\tau}^t-\hat{B}^t.
\]
Algorithm \ref{algo:1} summarizes the procedure that produces our proposed rematching estimators. An extra bias correction step may be needed depending on the data.

\begin{algorithm}[t]
\caption{Matching-and-Rematching Algorithm}
\label{algo:1}
  \KwData{$Z=\{(Y_i,D_i,X_i)\}_{i=1}^N$ and $M_0$.}
\KwResult{$\hat{\tau}^t$ and $\hat{\tau}$.}

Step 1 (Matching step): For {$i \in \{1,\ldots,N\}$}, perform $M_0$-nearest neighbour search to find $\mathcal{J}_{M_0}(i)$ defined in (\ref{eqt:JMF}).

Step 2 (Weighting step): For {$i \in \{1,\ldots,N\}$},  compute the number of times unit $i$ is matched, i.e., $K_{M_0}(i)$ defined in (\ref{eqt:KM0}).

Step 3 (Rematching step): For {$i \in \{1,\ldots,N\}$}, find the index of the nearest neighbour for potentially unmatched unit $i$ as in (\ref{eqt:JU}).

Step 4 (Re-weighting step): For {$\ell \in \{1,\ldots,N\}$}, obtain the number of times unit $\ell$ is matched to unmatched units as in (\ref{eqt:KU}). For {$i \in \{1,\ldots,N\}$}, obtain $K(i)$ defined in (\ref{eqt:KM}).

Step 5: Estimate $\tau$ and $\tau^t$ by rematching estimators defined in (\ref{eqt:tauHat}) and (\ref{eqt:tauTHat}).
\end{algorithm}

\section{Asymptotic properties}\label{sec:asy}
\subsection{Decomposition and bias analysis}\label{sec:bias}
Let $(Y_i, X_i, D_i)$, $i=1,\ldots, N$, be independent and identical copies
of $(Y, X, D)$.
Denote $\sigma^2(d, x) =\var(Y\mid D = d, X = x).$
Also denote the residual as $\epsilon_i=Y_i-\mu(D_i, X_i)$ for each $i$.
We first decompose $\hat{\tau}-\tau$ as
\begin{equation}\label{eqt:decomposition}
    \hat{\tau}-\tau =  \left\{\bar{\tau}(X) - \tau\right\} + R + B ,
\end{equation}
where
\begin{align*}
    \bar{\tau}(X) &= \frac{1}{N}\sum^N_{i=1}\{\mu(1,X_i) - \mu(0,X_i)\},\\
    R
    &= \frac{1}{N}\sum^N_{i=1} (2D_i - 1)\left\{1 +\frac{K(i)}{M_0}\right\}\epsilon_i,\\
    B
    &=  \frac{1}{N}\sum^N_{i=1}(2D_i - 1)\left[ \frac{1}{M(i) }\sum_{j\in \mathcal{J}(i)} \left\{\mu(1-D_i,X_i) - \mu\left(1-D_i,X_{j}\right)\right\}\right].
\end{align*}
Here, $\bar{\tau}(X)$
is the average conditional treatment effect,
$R $ is a weighted average of the residuals,
and $B $ is the conditional bias relative to $\tau(X)$.
The decomposition (\ref{eqt:decomposition}) is also employed by \cite{abadie2006large}. Similarly, we can decompose $\hat{\tau}^t-\tau^t$ as
\begin{equation}
    \hat{\tau} ^t -\tau^t =  \left\{\bar{\tau}^{t}(X) - \tau^t\right\} + R ^t + B ^t,\nonumber
\end{equation}
where
\begin{align*}
    \bar{\tau}^{t}(X)
    &= \frac{1}{N_1}\sum^N_{i=1}D_i\{\mu(1,X_i) - \mu(0,X_i)\}, \\
    R ^t
    &= \frac{1}{N_1}\sum^N_{i=1} \left\{D_i - (1-D_i)\frac{K(i)}{M_0}\right\}\epsilon_i, \\
    B ^t
    &=  \frac{1}{N_1}\sum^N_{i=1}D_i\left[ \frac{1}{M(i) }\sum_{j\in \mathcal{J}(i)} \left\{\mu(0,X_i) - \mu\left(0,X_{j}\right)\right\}\right].
\end{align*}
All asymptotic results are stated as $N\rightarrow\infty$ unless otherwise stated.
Regularity conditions for our asymptotic theory are included in the supplement;
see Assumptions {\ref{a:1}}--{\ref{as:3}}.

Let $\mathcal{X} \subset \mathbb{R}^k$ be the support of $X$ and  $\mathcal{X}_0 \subset \mathbb{R}^k$ be the support of $X$ given $D=0$. We establish bounds on the stochastic order of the conditional bias terms as follows.

\begin{theorem}[Conditional Bias]\label{thm:1}
Suppose Assumption {\ref{a:1}} holds.
\begin{enumerate}
    \item If Assumption {\ref{a23}} holds and $x\mapsto\mu(d,x)$ $(d=0,1)$ is Lipschitz continuous on $\mathcal{X}$, then $B  = O_p(N^{-1/k})$.
    \item If Assumption {\ref{a2p3p}} holds and  $x\mapsto\mu(0,x)$ is Lipschitz continuous on $\mathcal{X}_0$, then $B ^t = O_p(N_0^{-1/k})$.
\end{enumerate}
\end{theorem}
The theorem above shows that the bias of our proposed rematching estimators is of the same order as that of simple matching estimators, which makes the causal interpretation of our proposals as credible as that of simple matching estimators.
Moreover, Theorem \ref{thm:1} has a great impact on the asymptotic normality of rematching estimators.

From part 1 of the theorem,
$B= O_p(N^{-1})$ when the dimension of the continuous covariate is $k = 1$.
As we shall see in Section \ref{sec:normality},
$N^{1/2}\{\Bar{\tau}(X) - \tau\} = N^{1/2}R = O_p(1)$ is asymptotically normal
under regularity conditions.
Consequently,
the asymptotic normality of $N^{1/2}(\hat{\tau} - \tau)$ is achieved, hence,
the rematching estimator is $N^{1/2}$-consistent.
When there are more than one continuously distributed covariate, i.e., $k>1$,
bias correction is needed in order to attain the $N^{1/2}$-consistency
of the rematching estimator.

Part 2 of this theorem is particularly useful since matching estimators nearly always estimate the average effect for the treated when a large reservoir of controls is available.
Generally, the bias is ignorable when there is only one continuous covariate or the number of controls is sufficiently large.
To be precise,
if the two group sizes go to infinity at different rates such that $N_1 = O_{p}(N^s_0)$,
then $B^t = O_p\{N^{1/(sk)}_1\}$.
Therefore, if $s < 2/k$, then $B^t = o_p(N^{-1/2}_1)$, which will be dominated in the large sample distribution of $\hat{\tau}-\tau$ by $\bar{\tau}^{t}(X) - \tau$ and $R^t$, which are of size $O_p(N^{-1/2}_1)$.

\subsection{Variance improvement}
We first investigate the conditional variance of $\hat{\tau}$.
Conditional on $X_{1:N}=(X_1, \ldots, X_N)^{\T}$ and $D_{1:N}=(D_1, \ldots, D_N)^{\T}$,
the weighted version of the number of times a unit is used as a match, $K(i)$, is deterministic.
Hence, according to (\ref{eqt:tauHat}), the conditional variance of $\hat{\tau} $ is
\begin{equation*}
    \var(\hat{\tau} \mid D_{1:N},X_{1:N})
        = \frac{1}{N^2}\sum^N_{i=1}\left\{1+\frac{K(i)}{M_0}\right\}^2
        \sigma^2(D_i,X_i).
\end{equation*}
Similarly, according to (\ref{eqt:tauTHat}), its counterpart for the average effect of the treatment on the treated is
\begin{equation*}
    \var(\hat{\tau}^t \mid D_{1:N},X_{1:N})
        = \frac{1}{N_1^2}\sum^N_{i=1}\left\{D_i-(1-D_i)\frac{K(i)}{M_0}\right\}^2
        \sigma^2(D_i,X_i).
\end{equation*}
Let $V^R  = N\var(\hat{\tau} \mid D_{1:N},X_{1:N})$ and $V^{R,t}  = N_1\var(\hat{\tau}^t \mid D_{1:N},X_{1:N})$.
The lemma below shows the finiteness of the expectation of these conditional variances.
\begin{lemma}[Conditional variances]\label{lem:1}
Suppose Assumption {\ref{a:1}} holds.
\begin{enumerate}[ref=\ref{lem:1}.\arabic*]
    \item \label{lem:1.1}Suppose Assumption {\ref{a23}} holds.
        For $i \in \{1,\ldots,N\}$,
        $K(i) = O_p(1)$, and $\E\{K_M^q(i)\}$ is bounded uniformly in $N$ for any $q>0$.
    \item \label{lem:1.2}Suppose Assumption {\ref{a2p3p}} holds.
        For $i \in \{1,\ldots,N\}$, $(N_0/N_1)\E\{K^q_M(i)\mid D_i=0\}$ is bounded uniformly in $N$ for any $q>0$.
    \item Suppose $x\mapsto\sigma^2(d,x)$ is Lipschitz continuous in $\mathcal{X}$,
        for $d = 0,1$. If Assumption {\ref{a23}} holds, then
        $
            \E(V^R)=O(1).
        $
        If Assumption {\ref{a2p3p}} holds, then $
            \E(V^{R,t})=O(1).
        $
\end{enumerate}
\end{lemma}

According to Theorem \ref{thm:1}, $B=O_p(1/N)$ and $B^t = O_p(1/N_0)$ when $k=1$. And $B^t = o_p(N^{-1/2}_1)$ when the number of controls is sufficiently large compared to the number of treated.
In these cases, the bias terms $B$ and $B^t$ are dominated in the large sample distributions of $\hat{\tau}-\tau$ and $\hat{\tau}^t-\tau^t$, respectively, and given the value of $N_1$,
the unconditional variances of $\hat{\tau}$ and $\hat{\tau}^t$ are
\[
    \var(\hat{\tau} ) \sim \frac{\E(V^R) + V^{\tau(X)}}{N},
    \quad
    \var(\hat{\tau}^t) \sim \frac{\E(V^{R,t}) + V^{\tau(X),t}}{N_1},
\]
respectively, where $a_N \sim b_N$ means $a_N/b_N\rightarrow 1$ as $N\rightarrow\infty$; and
\begin{align*}
    V^{\tau(X)} &= \E\left\{  \mu(1, X) - \mu(0, X) - \tau \right\}^2,\\
    V^{\tau(X),t} &= \E\left[ \left\{ \mu(1, X) - \mu(0, X) - \tau^t\right\}^2\mid D=1 \right].
\end{align*}

According to \cite{abadie2006large}, investigating the asymptotic variance of matching estimators when $k \ge 2$ is challenging.
We leave the general case for future study.
The theorem below shows the asymptotic efficiency gain for the special case with a scalar covariate, i.e., $k = 1$.
\begin{theorem}[Asymptotic efficiency gain]\label{thm:generalVar}
    Suppose $k=1$. If Assumptions {\ref{a:1}}, {\ref{a23}} and {\ref{a:4}} hold, and $f_0(x)$ and $f_1(x)$ are continuous on $\mathcal{X}$, then
    \begin{align*}
        N\var(\hat{\tau} )
    &\leq N\var(\hat{\tau}_0 )\\ &\qquad=
    \left(1+\frac{1}{2 M}\right) \E\left[\frac{\sigma_{1}^{2}(X_{i})}{e(X_{i})}+\frac{\sigma_{0}^{2}(X_{i})}{1-e(X_{i})}\right] + V^{\tau(X)} \\
&\qquad\qquad -\frac{1}{2 M} \E[e(X_{i}) \sigma_{1}^{2}(X_{i})+(1-e(X_{i})) \sigma_{0}^{2}(X_{i})] +o(1),
    \end{align*}
where $e(x) = \pr(D = 1|X = x)$ is the propensity score.
\end{theorem}
The strict inequality holds when $K_{M_0}(i)=0$ for some $i$, i.e., when rematching can be done.
The exact form of $N\var(\hat{\tau}_0 )$ in the theorem above is given by \cite{abadie2006large}.
The asymptotic unconditional variance of $\hat{\tau}$ is not available due to the complicated form of $K(i)$; see Remark \ref{rem:exactForm}.

\begin{remark}\label{rem:exactForm}
    Unlike $K_{M_0}(i)$, which has a simple form and follows a binomial distribution, our $K(i)$ has a more complicated form, making it  difficult to derive the unconditional variance of the estimator.
    Specifically, we need to consider the distribution and the first two moments of $X/(Y+M_0)$, where $X$ and $Y$ are two dependent binomial random variables.
    It is difficult to find moments of the ratio of two binomial distributions, not to mention rewriting the moments in terms of propensity scores like what \cite{abadie2006large} and \cite{hahn1998role} did.
    Thus, it is hard to derive explicitly
    the exact efficiency loss relative to the semi-parametric efficiency bound in \cite{hahn1998role} or the exact efficiency gain relative to the standard matching estimator in \cite{abadie2006large}.
\end{remark}

To derive a representation of the unconditional variance, we need the following notation.
Let $\Bin(n, p)$ denote the binomial distribution with $n \in \mathbb {N}$ trails and success probability $p \in [0,1]$.
Let $\Bern(p)$ denote the Bernoulli distribution with success probability $p$.
Now, we let $p=\pr(D=1)$ denoting the treated ratio and let $f_d(x)$ denote the density function of $X$ under $D=d$.
The theorem below gives the form of unconditional variances.

\begin{theorem}[Unconditional variances]\label{thm:4}
Suppose that $k = 1$,
$f_0(x) = f_1(x)$,
and $M_0N_1<N_0$.
Further suppose that $x \mapsto \sigma^2(d,x) =: \sigma^2_d$ is a constant as a function of $x$ for each $d\in\{0,1\}$.
Define the random variables $K$, $I$, and $H$ as follows given the values of $N_0$ and $N_1$:
\begin{align*}
    K &\sim\Bin(M_0N_0,1/N_1), \quad
    I \sim \Bern(M_0N_1/N_0), \quad\\
    [H \mid I]&\sim \Bin(N_0-M_0N_1+I-1,1/N_1).
\end{align*}
\begin{enumerate}
    \item If Assumptions {\ref{a23}} and {\ref{a:4}} hold,
            then
            \begin{align*}
                N\var(\hat{\tau} )
                    &\rightarrow   \sigma^2_1 p    \E\left(1+\frac{K}{M_0}\right)^2 \\
                       &\quad+ \sigma^2_0 (1-p)\E\left( 1+\frac{I}{H+M_0}
                                                +\frac{1-I}{H+1+M_0}\right)^2
                       + V^{\tau(X)} ,\\
                N\var(\hat{\tau}_{0} )
                    &\rightarrow   \sigma^2_1 p    \E\left(1+\frac{K}{M_0}\right)^2
                        + \sigma^2_0 (1-p)\E\left(1+\frac{I}{M_0}\right)^2
                        + V^{\tau(X)} .
            \end{align*}
    \item If Assumptions {\ref{a2p3p}} and {\ref{a:4}} hold, then
            \begin{align*}
                N_1\var(\hat{\tau}^{t})
                    &\rightarrow\sigma^2_1p + \sigma^2_0(1-p) \E\left(\frac{I}{H+M_0} +\frac{1-I}{H+1+M_0}\right)^2
                     +V^{\tau(X),t} , \\
            N_1\var(\hat{\tau}_{0}^{t})
                &\rightarrow\sigma^2_1p + \sigma^2_0(1-p)\E\left(\frac{I}{M_0}\right)^2+V^{\tau(X),t}.
            \end{align*}
\end{enumerate}

\end{theorem}
Note that $K$ and $[K_{M_0}(i)\mid D_i=1]$ have identical distributions,
and so do $I$ and $[K_{M_0}(i)\mid D_i=0]$. Also,
$[H\mid I]$, $M_{\re}(\ell)$, and $M_{\re}(\mathcal{J}_{re}(i))-1$ follow the same distribution.
To study the variance improvement of our proposed estimators over simple matching estimators under this assumption,
we need the condition that $M_0N_1<N_0$ to simulate scenarios with unmatched controls.
When there are no unmatched controls after the simple matching,
the rematching estimator $\hat{\tau}^t$ and the simple matching estimator $\hat{\tau}_0^t$ produce the same estimate.
According to the proposition below, our proposal has a uniform variance improvement over the simple matching estimator.

\begin{proposition}\label{prop:3}
Under the conditions in Theorem \ref{thm:4},
the difference between $N_1\var(\hat{\tau}_{0}^{t})$ and $N_1\var(\hat{\tau}^{t})$
satisfies that
\begin{align*}
    &N_1\left\{\var(\hat{\tau}_{0}^{t})-\var(\hat{\tau}^{t})\right\}
         \\&\qquad\rightarrow
         \sigma_0^2(1-p) \E\left\{\left(\frac{I}{M_0}\right)^2
                            - \left( \frac{I}{H+M_0}+\frac{1-I}{H+1+M_0}\right)^2\right\}>0.
\end{align*}
\end{proposition}

Based on Theorem \ref{thm:4},
we compare rematching estimators with simple matching estimators by simulation experiments.
Since matching estimators are used nearly always when the quantity of interest is the average effect on the treated and there are more controls than treated, we only include the results
for the estimators of $\tau^t$ and consider a design with the treated ratio $p=0.05, 0.1,0.15,0.2,0.25,0.3$, and $M_0=1,2,3,4,5.$
For each case, we set the sample size as $N = 100$ and the conditional error variance as $\sigma^2(d, x)=1$.
Assume we have prior knowledge that the average treatment effect conditional on the covariates is a constant, i.e., $\mu(1,x) - \mu(0,x) = \tau^t$ for all $x$, and hence, $V^{\tau(x),t} = 0$.
We use the percentage decrease in variance, i.e., $\{\var(\hat{\tau}_{0}^{t})-\var(\hat{\tau}^{t})\}/\var(\hat{\tau}_{0}^{t})$,
as a measure of the ability of our rematching process to gain efficiency.

\begin{figure}[t!]
	\begin{center}
		\includegraphics[width=.9\textwidth]{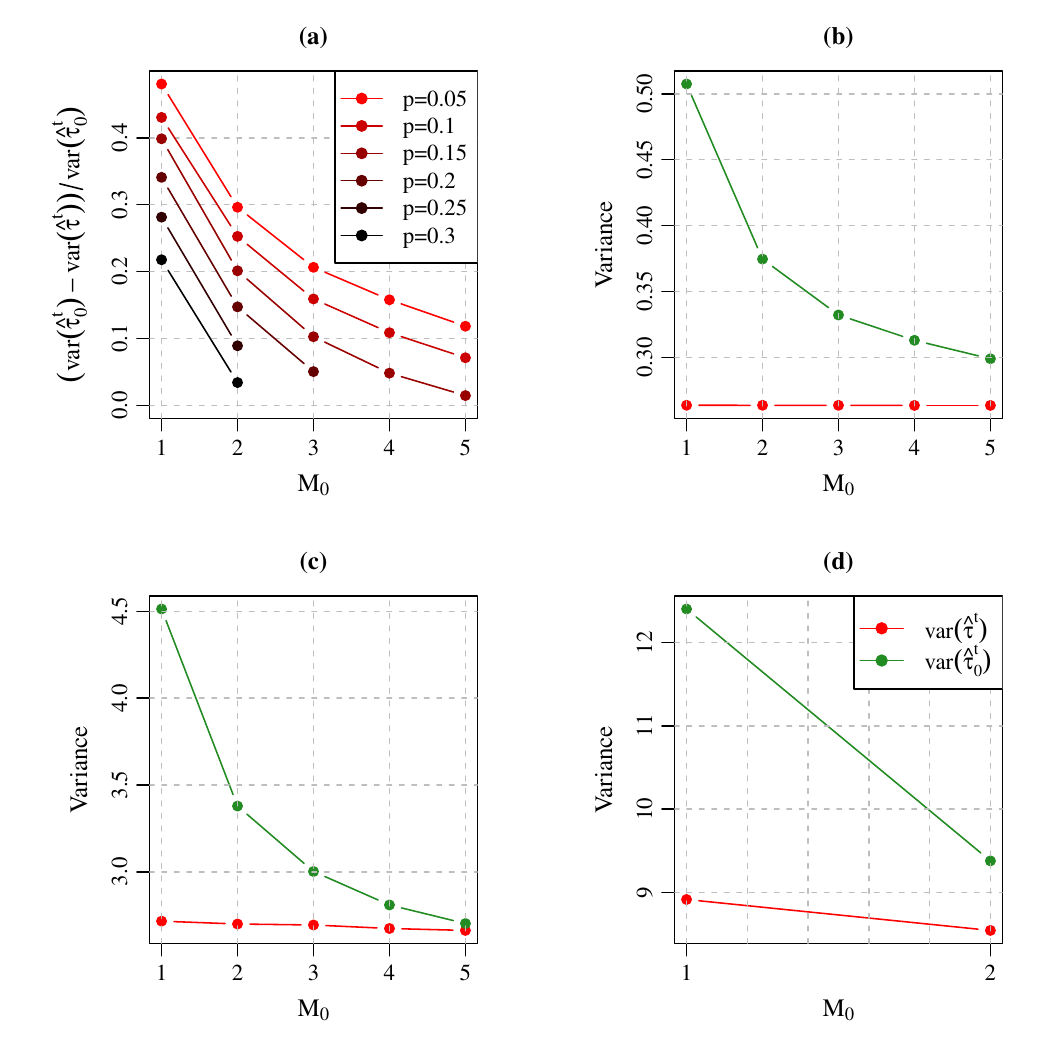}
		\caption{Plots of simulation results of variances of $\hat{\tau}_{0}^{t}$ and $\hat{\tau}^{t}$ across different $M_0$ in various settings: (a) Plot of percentage variance reduction with $p=0.05, 0.1,0.15,0.2,0.25,0.3$; (b)--(d) Variance plots of $\hat{\tau}_{0}^{t}$ (green circles) and $\hat{\tau}^{t}$ (red circles) fixing $p$ at $0.05,0.15, 0.25$, respectively.
The variances are approximated by simulation based on $2^{14}$ replications.}
		\label{fig:uncondVar}
	\end{center}
\end{figure}

Figure \ref{fig:uncondVar} visualizes the  reduction in variance across
different treated ratios and different numbers of fixed matches.
Since Theorem \ref{thm:4} assumes there are unmatched units after the simple matching, i.e., $M_0N_1<N_0$,
the maximum possible $M_0$ depends on the treated ratio $p = N_1/(N_0+N_1)$.
In any case, the proposed estimator $\hat{\tau}^t$ is always more precise than the traditional $\hat{\tau}_0^t$.
Fixing $M_0$ at a particular value, the smaller the treated ratio,
the greater the percentage reduction in variance.
Given a specific $p$, the smaller the number of fixed matches,
the greater the variance reduction percentage.
The results with $M_0=1$ are worth noticing as people typically fix $M_0$ at one when using matching estimators.
In particular, when $M_0=1$ and $p=0.05$, the variance is reduced by around 50\%;
see also Remark \ref{rem:incM0}.
Although previous results are based on the condition that $f_0(x)=f_1(x)$,
it is interesting to discuss the impact of the density ratio on the variance reduction; see Remark \ref{rem:denRatio}.

\begin{remark}\label{rem:incM0}
    Since $M_0<N_0/N_1$ and $f_0(x)=f_1(x)$,
    the variance of the rematching estimator $\hat{\tau}^t$ is roughly a constant across $M_0$.
    To reach roughly the same variance as the rematching estimator,
    we need to increase $M_0$ to the largest value that satisfies $M_0<N_0/N_1$ as specified in Theorem \ref{thm:4}, i.e., $M_0 < (1-p)/p$.
However, even if $M_0$ is increased to the largest possible value, the variance of the matching estimator is still larger than the variance of the rematching estimator.
\end{remark}

\begin{remark}\label{rem:denRatio}
    Intuitively, if the density ratio is large across different $X$, i.e.,
    the covariate distribution is quite different between the treated and the control groups,
    our estimator with rematching will utilize more information than the matching estimator when a small to moderate $M_0$ is chosen.
    Because when $f_0(x)/f_1(x)$ is large, the number of neighbours within a particular distance is quite different among units.
    Then, our estimator with a small $M_0$ is more efficient than the matching estimator with a large but not large enough $M_0$
    since a really large $M_0$ is needed to include more information in this case.
    But increasing $M_0$ infinitely is risky because additional bias may be introduced.
    Moreover, the higher the density ratio, the more pronounced the bias caused by the increase in $M_0$ for the sake of efficiency.
\end{remark}

\subsection{Consistency and asymptotic normality}\label{sec:normality}
Here, we show that the proposed matching estimators are consistent for $\tau$ and $\tau^t$. Without the bias term or after the bias correction, they are $N^{1/2}$-consistent and asymptotically normal;
see Theorems \ref{thm:2}, \ref{thm:3}, and \ref{thm:5}.

Let $\Normal(\mu,\sigma^2)$ denote the normal distribution with mean $\mu\in\mathbb{R}$ and variance $\sigma^2\in \mathbb{R}^+$.
\begin{theorem}[Consistency]\label{thm:2}
Suppose Assumption {\ref{a:1}} holds.
\begin{enumerate}
\item If Assumptions {\ref{a23}} and {\ref{a:4i}} hold, then $\hat{\tau}  - \tau \rightarrow  0$ in probability.
\item If Assumptions {\ref{a2p3p}} and {\ref{a:4i}} hold, then $\hat{\tau}^t  - \tau^t \rightarrow 0$ in probability.
\end{enumerate}
\end{theorem}

\begin{theorem}[Asymptotic Normality]\label{thm:3}
Suppose Assumptions {\ref{a:1}} and {\ref{a:4}} hold.
\begin{enumerate}
    \item If Assumption {\ref{a23}} holds, then
            $
                \{V^R  + V^{\tau(X)}\}^{-1/2}{N}^{1/2}(\hat{\tau}  - B  - \tau)
                \rightarrow \Normal(0, 1)
            $
            in distribution.
    \item  If Assumption {\ref{a2p3p}}  holds, then
            $
                \{V ^{R,t} + V^{\tau(X),t}\}^{-1/2}N_1^{1/2}(\hat{\tau}^t  - B^t  - \tau^t) \rightarrow  \Normal(0, 1)
            $
            in distribution.
\end{enumerate}
\end{theorem}
Theorem \ref{thm:3}, which adopts a similar form as Theorem 4 in \cite{abadie2006large}, states the asymptotic normality of $\hat{\tau}$ and $\hat{\tau}^t $ after subtracting $B$ and $B^t$.
Though a similar form is adopted, the formulas of some terms including $V^R$, $V^{R,t}$, $B$, and $B^t$ adopt different forms from those in \citet{abadie2006large} since our matching-and-rematching procedure can produce matching results that differ greatly from simple matching results.

\begin{theorem}[Consistency and Asymptotic Normality for $\hat{\tau}_{\bc}$ and $\hat{\tau}^t_{\bc}$]\label{thm:5}
Suppose that Assumptions {\ref{a:1}} and {\ref{a:4}}--{\ref{as:3}} hold.
\begin{enumerate}
    \item Suppose Assumption {\ref{a23}} hold, then
            $N^{1/2}(B-\hat{B})\rightarrow0$ in probability, and
            \[
                \{V^R  + V^{\tau(X)}\}^{-1/2}N^{1/2}(\hat{\tau}_{\bc}  - \tau)
                \rightarrow  \Normal(0, 1)
\]
        in distribution.
    \item Suppose Assumption {\ref{a2p3p}} hold, then
        $N_1^{1/2}(B^t-\hat{B}^t)\rightarrow 0$ in probability,
        and
        \[
            \{V^{R,t}  + V^{\tau(X),t}\}^{-1/2}N_1^{1/2}(\hat{\tau}^t_{\bc}  - \tau^t)
            \rightarrow  \Normal(0, 1)
\]
        in distribution.
\end{enumerate}
\end{theorem}

The theorem above implies that we can estimate the bias of rematching estimators
at a similar speed to that of estimating the bias of simple matching estimators,
which is faster than $N^{1/2}$ for the estimated average treatment effect and
$N_1^{1/2}$ for the estimated average treatment effect on the treated.
Also, the normalized variance remains the same after the bias correction.

\subsection{Variance estimation}\label{sec:varEst}
By the matching strategy mentioned in \cite{abadie2006large}, we  estimate $\sigma^2(D_i, X_i)$ as
\[
    \hat{\sigma}^2(D_i, X_i) = \frac{J}{J + 1}
    \left( Y_i - \frac{1}{J}\sum_{j=1}^J Y_{l_j (i) }\right)^2 ,
\]
where $J$ is the number of matches used, $l_j (i)$ is the index of the $j$th closest match of the same treatment group for unit $i$,
$\ell_J(i) = \{l_1(i),\ldots, l_J(i)\}$, and
{\footnotesize
\begin{align*}
\ell_J(i) = \left\{j \in  \{1,..., N\} : D_j = D_i \text{ and }
\sum_{l\neq j:D_l=D_i} \mathbb{1}\{\lVert X_l- X_i\rVert  \leq
 \lVert X_j - X_i \rVert \} \leq J
\right\}
.
\end{align*}}
Then, we propose to estimate $V=V^R  + V^{\tau(X)}$ and $V^t=V^{R,t}  + V^{\tau(X),t}$ by
\begin{align*}
    \hat{V}
    &= \frac{1}{N}\sum_{i=1}^N
        \{\Tilde{Y}_i(1)-\Tilde{Y}_i(0)-\hat{\tau}\}^2\\
        &\qquad\qquad+ \frac{1}{N}\sum_{i=1}^N  \left[\left\{\frac{K(i)}{M_0}\right\}^2+\left(\frac{2M_0-1}{M_0}\right)\left\{\frac{K(i)}{M_0}\right\}\right] \hat{\sigma}^2(D_i, X_i) ,  \\
    \hat{V}^t
    &= \frac{1}{N_1}\sum_{\substack{1\le i\le N \\ D_i=1}}
        \{Y_i-\Tilde{Y}_i(0)-\hat{\tau}^t\}^2\\
        &\qquad\qquad+ \frac{1}{N_1}\sum_{i=1}^N
        (1-D_i)\left[\frac{K(i)\{K(i)-1\}}{M_0^2}\right]
        \hat{\sigma}^2(D_i, X_i) ,
\end{align*}
respectively.
The consistency is guaranteed as follows, which allows us to perform statistical inference on $\tau$ and $\tau^t$.
For example, confidence interval for $\tau$ and $\tau^t$ can be computed.

\begin{theorem}\label{thm:6}
If Assumptions {\ref{a:1}}, {\ref{a23}} and {\ref{a:4}} hold, then $|\hat{V}-V|=o_p (1)$. If Assumptions {\ref{a:1}}, {\ref{a2p3p}} and {\ref{a:4}} hold, then $|\hat{V}^t-V^t|=o_p(1)$.
\end{theorem}

\section{Simulation experiments}\label{sec:sim}
\begin{figure}[t!]
	\begin{center}
		\includegraphics[width=\textwidth]{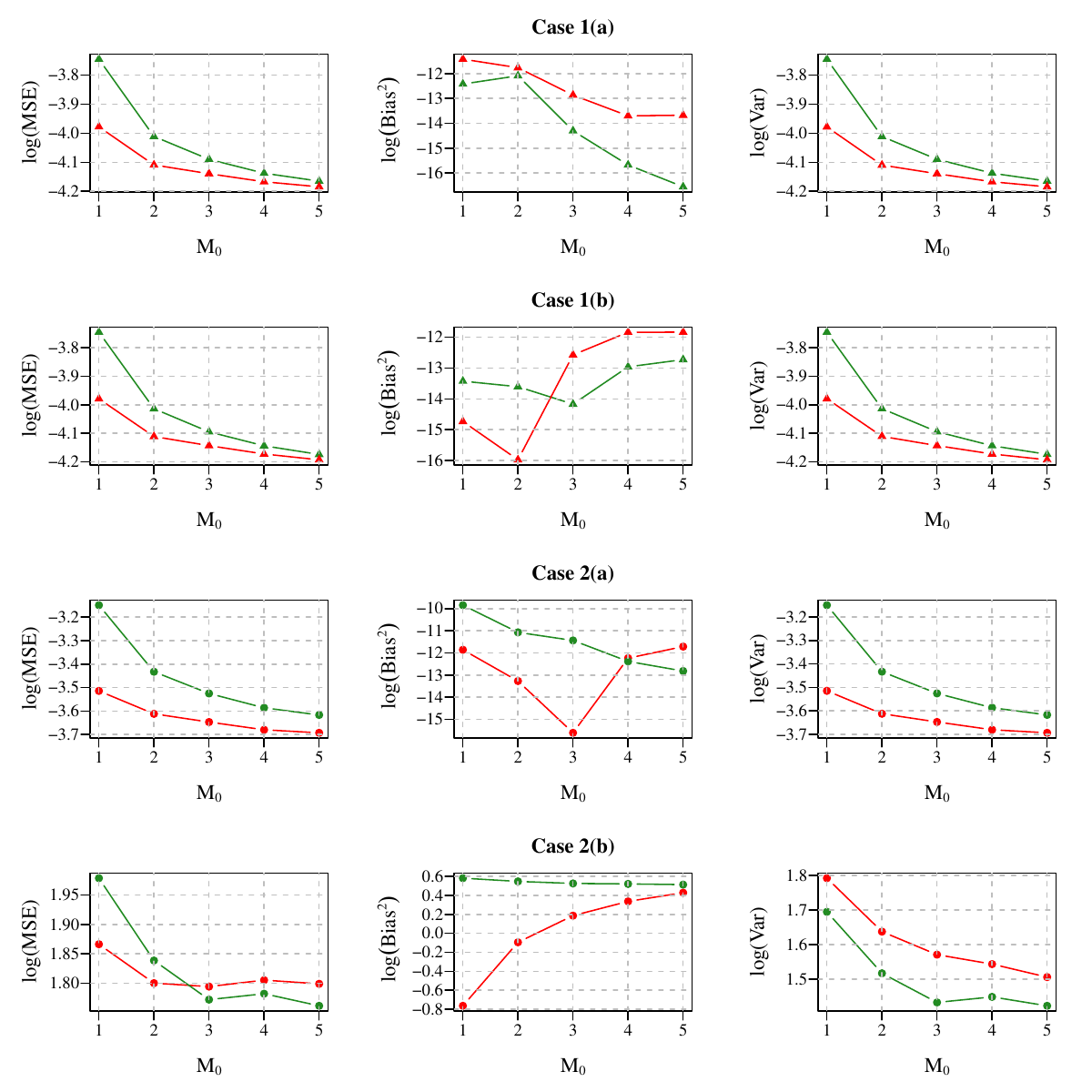}
		\caption{A graph showing the performance comparison between $\hat{\tau}_{0}^{t}$ (green triangles) and $\hat{\tau}^{t}$ (red triangles) when $X$ is a scalar and between their bias-corrected versions, $\hat{\tau}^{t}_{0,\bc}$ (green circles) and $\hat{\tau}^{t}_{\bc}$ (red circles), when $X$ is multi-dimensional.}
		\label{fig:sim_scalar_multi}
	\end{center}
\end{figure}

We evaluate the performance of rematching estimators by Monte Carlo simulation with ten design cases:
Cases 1(a)--(e) concern a scalar covariate, whereas
Cases 2(a)--(e) concern multiple covariates.
Due to space constraints,
only four cases are shown in Figure \ref{fig:sim_scalar_multi} with the complete results deferred to the supplement.
We follow previous studies and focus on the average treatment effect on the treated $\tau^t$.
Based on \cite{frolich2004finite}, \cite{busso2014new} and \cite{otsu2017bootstrap}, we consider the following data-generating process: For $i=1, \ldots, N$ and $d=0,1$,
\begin{align*}
    Y_i(d) &= \tau d  + m(X_i) + \epsilon_i,
    \quad
    D_i = \mathbb{1}\left\{P(X_i) \geq \upsilon_i\right\}\mathbb{1}(\xi_i>c),
\end{align*}
where
$\epsilon_i \sim \Normal(0, 0.5^2)$,
$\upsilon_i\sim \Beta(1, 1)$, and
$\xi_i\sim \Beta(1, 1)$ are mutually independent;
$x \mapsto P(x)$ is a function for specifying the propensity score;
$x \mapsto m(x)$ is a mean function; and
$c$ is a constant.
Here $\Beta(\alpha, \beta)$ denotes the beta distribution with parameters $\alpha, \beta>0$.
For Cases 1(a)--(e), let $X_i\sim \Beta(1.2, 1.2)$.
For Cases 2(a)--(e), let $X_i$ be a 6-dimensional covariate containing
both continuous and discrete components.
The detailed simulation designs,
including the settings of $P(\cdot)$, $m(\cdot)$, $c$, and the distribution of $X_i$,
are available in the supplement.

We first consider cases with a scalar $X$.
The bias term is negligible since $B^t = O_p(1/N_0)$ is of sufficiently low order.
From Figure \ref{fig:sim_scalar_multi}, we find that our proposed estimator $\hat{\tau}^t$ performs better than the simple matching estimator $\hat{\tau}_0^t$ across different choices of $M_0$ in all scalar covariate cases,
i.e., Cases 1(a) and 1(b), as the rematching process allows more efficiency gain while maintaining a negligible bias when $X$ is a scalar.
Also, the mean squared error
is comprised mainly of the variance,
which makes the leftmost and the rightmost graphs almost identical in each case.
The gain in precision by increasing $M_0$ diminishes rapidly after 4, which is consistent with the findings of \cite{rosenbaum2020modern}.
The declining trend in the mean squared error ($\MSE$) and variance, along with the convergence of both methods as $M_0$ increases, may give the impression that the rematching process is redundant as higher $M_0$ already ensures a more precise estimation by $\hat{\tau}_0^t$.
However, as we shall see later, the bias can increase to a non-negligible magnitude with $M_0$ when there are multiple covariates.
This is undesirable given the focus on decreasing the bias in the literature of causal inference \citep{imbens2009recent}.

For all multivariate cases, we compare the performance of the rematching estimator, the simple matching estimator, and their bias-corrected versions.
We also include in the comparison the genetic matching estimator \citep{diamond2013genetic}, which uses a state-of-art iterative algorithm to maximize a criterion related to covariate balance.
Since the mean squared errors of $\hat{\tau}^t_{\bc}$ and $\hat{\tau}^{t}_{0,\bc}$ are much lower compared with others, we move the complete simulation result to the supplement and show only the results for $\hat{\tau}^t_{\bc}$ and $\hat{\tau}^{t}_{0,\bc}$ in Figure \ref{fig:sim_scalar_multi}.
The performance of the bias-corrected estimators depends on the regression-based bias estimation.
When the deviation from the true outcome model of the specified model is small, Case 2(a) for instance,
it is favourable to increase $M_0$ for efficiency,
which is consistent with the findings in \cite{lin2023estimation},
meaning that we can increase the number of fixed matches for $\hat{\tau}^t_{0,\bc}$ with the sample size for better estimation.
However, using our proposed estimator gives an ideal result even at a very small number of fixed matches (e.g., $M_0=1$).
Also, we may not always be able to specify the model correctly in practice.
In Case 2(b), where the bias is not well-corrected, an interesting finding is that our proposal $\hat{\tau}_{\bc}$ is able to achieve a lower bias across $M_0$.
Considering that bias is a good measure of the covariate balance and a smaller bias means a more valid causal interpretation, our proposal with a small $M_0$ is generally better than the original proposal.

In all cases, our matching-and-rematching procedure gains efficiency while maintaining low bias, which sits well with the focus on the credibility of the causal inference in the literature \citep{imbens2009recent}.
Practically, people simply fix
$M_0=1$
when using matching estimators.
For this particular choice of $M_0$, our matching-and-rematching procedure gives better estimation results since it allows more efficiency gain when the bias is negligible or the outcome model is well-specified and less precision loss when the outcome model is misspecified.

\section{Empirical application}\label{sec:app}
We apply rematching estimators to analyze the National Supported Work data,
an evaluation of a job training program analyzed by \cite{lalonde1986evaluating}, \cite{heckman1989choosing}, \cite{dehejia1999causal}, \cite{imbens2003sensitivity}, and \cite{smith2005does}.
The dataset is available on Rajeev Dehejia's website
(\texttt{http://users.nber.org/\~{}rdehejia/nswdata2.html}).
It contains experimental and non-experimental samples;
see the supplement for details.
As is common in previous studies, we focus on the average effect of the program on earnings for the treated.

\begin{table}[t!]
\caption{Experimental and nonexperimental estimates for the NSW data.}
  \centering
  \renewcommand{\arraystretch}{1.1}
  \setlength{\tabcolsep}{3.0pt}
  \begin{tabular}{|lcccccccccc|}\hline
    & \multicolumn{2}{c}{$M_0 = 1$}
    & \multicolumn{2}{c}{$M_0 = 4$}
    & \multicolumn{2}{c}{$M_0 = 16$}
    & \multicolumn{2}{c}{$M_0 = 64$}
    & \multicolumn{2}{c|}{$M_0 = N_0$} \\
    & \multicolumn{1}{c}{Est.} & \multicolumn{1}{c}{(SE)} & \multicolumn{1}{c}{Est.} & \multicolumn{1}{c}{(SE)}& \multicolumn{1}{c}{Est.} & \multicolumn{1}{c}{(SE)}& \multicolumn{1}{c}{Est.} & \multicolumn{1}{c}{(SE)}& \multicolumn{1}{c}{Est.} & \multicolumn{1}{c|}{(SE)} \\\hline
    \multicolumn{11}{|l|}{Experimental} \\
    $\quad$ $\hat{\tau}_0^{t}$  & 1.82 & (0.85) & 2.08 & (0.72) & 1.97 & (0.68) & 2.23 & (0.69) & 1.79 & (0.67)\\
    $\quad$  $\hat{\tau}^{t}_{0,\bc}$  & 1.87 & (0.84) & 1.90 & (0.73) & 1.74 & (0.68) & 1.63 & (0.69) & 1.77 & (0.66) \\
    $\quad$  $\hat{\tau}^{t}$  & 1.94 & (0.81) & 2.04 & (0.72) & 1.97 & (0.68) & 2.23 & (0.69) & 1.79 & (0.67) \\
    $\quad$  $\hat{\tau}^{t}_{\bc}$   & 1.99 & (0.80) & 1.88 & (0.72) & 1.75 & (0.68) & 1.63 & (0.69) & 1.77 & (0.66)\\\hline
    \multicolumn{11}{|l|}{Non-experimental} \\
    $\quad$ $\hat{\tau}_0^{t}$  & 1.14 & (1.25) & 1.23 & (1.08) & 0.66 & (1.02) & $-0.32$ & (0.86) & $-3.65$ & (0.75) \\
    $\quad$ $\hat{\tau}^{t}_{0,\bc}$  & 1.82 & (1.26) & 2.06 & (1.08) & 1.96 & (1.03) & 2.32 & (0.88) & 1.83 & (0.77) \\
    $\quad$ $\hat{\tau}^{t}$  & 1.05 & (1.22) & 1.13 & (1.08) & 0.61 & (1.02) & $-0.32$ & (0.86) & $-3.65$ & (0.75) \\
    $\quad$ $\hat{\tau}^{t}_{\bc}$  & 2.02 & (1.22) & 2.11 & (1.08) & 1.96 & (1.04) & 2.29 & (0.88) & 1.83 & (0.77)\\\hline
  \end{tabular}
  \label{tb:est}
\end{table}

Table \ref{tb:est} summarizes the results for the estimates and standard errors.
Although the true treatment effect is unknown,
the difference-in-mean estimate for the experimental data, which is 1.79 with a standard error of 0.67, can be regarded as a benchmark.
The 95\% confidence interval obtained by a normal approximation to the limiting distribution is $[0.48, 3.11]$.
For the non-experimental sample, the estimated treatment effects by our proposed bias-adjusted estimator and by the bias-adjusted simple matching estimator are all inside the experimental 95\% confidence interval.
It is found that the standard errors of our proposal $\hat{\tau}^t_{\bc}$ are smaller than those of the bias-adjusted matching estimator by \cite{abadie2011bias} when $M_0=1$, which is consistent with our goal of increasing efficiency by rematching.
However, because each pair of the treated and the control groups are close in size, which results in a small number of unmatched controls for large $M_0$,
the improvement in standard error is insignificant for $M_0>1$ according to Table \ref{tb:est}.
Moreover, setting $M_0=1$ results in the use of the highest quality matches,
reducing bias in both matching and rematching estimators.
Therefore, we consider $M_0=1$ for our analysis below.

In particular, we are interested in testing whether the treatment effect is positive, i.e.,
testing $H_0:\tau^t=0$ against $H_1: \tau^t>0$.
The $p$-value can be computed by the normal approximation to the asymptotic distribution.
Based on the non-experimental sample, we find that when $M_0=1$,
the $p$-values based on the classical matching estimator $\hat{\tau}^t_{0,\bc}$ and
our proposed rematching estimator $\hat{\tau}^t_{\bc}$ are
$1-\Phi(1.82/1.26)=7.43\%$ and $1-\Phi(2.02/1.22)=4.88\%$, respectively,
where $\Phi(\cdot)$ is the distribution function of $\Normal(0,1)$.
It means that our proposed test
successfully identifies the treatment effect at 5\% significance level
while $\hat{\tau}^t_{0,\bc}$ fails in doing so.

In a nutshell, our proposed estimator $\hat{\tau}^t_{\bc}$ is generally more efficient, resulting in a more powerful test and maintaining statistical validity.

\section{Conclusion}
In this article, we propose new matching estimators of treatment effects and derive their large sample properties.
In contrast to the simple matching, our matching-and-rematching procedure gains efficiency without increasing the bias too much by rematching for the treated units from an opposite direction, which matches each treated unit with a variable number of unmatched controls and increases the matched sample size.
Our method is applicable to both the average treatment effect and its counterpart for the treated population.
Simulation results indicate that our method works well in finite samples, suggesting it may be a useful estimator in practice.
Finally, an application to the National Supported Work data reveals an interesting test result.

\section*{Supplementary Material}
The supplement contains
technical assumptions,
proofs of main results,
additional simulation results,
a detailed description of the real-data application, and
a summary of existing work;
see Sections S1--S5, respectively, included below.

\par
\section*{Acknowledgments}
The authors would like to thank the anonymous referees, an Associate Editor, and the Editor for their constructive comments that improved the scope and theoretical development of the paper.
We are particularly grateful to the referees for providing valuable suggestions to improve the argument in the proof of Theorem \ref{thm:1}.

\section*{Funding}
Chan was partially supported by the Early Career Scheme 24306919 and the General Research Fund 14307922, provided by the Research Grants Council of HKSAR.
\par

\clearpage
\section*{Supplementary Material}
\addcontentsline{toc}{section}{Supplementary Material}
\setcounter{section}{0}
\setcounter{equation}{0}
\setcounter{figure}{0}
\setcounter{table}{0}
\setcounter{assumption}{0}
\setcounter{remark}{0}
\renewcommand{\thesection}{S\arabic{section}}
\renewcommand{\thefigure}{S\arabic{figure}}
\renewcommand{\thetable}{S\arabic{table}}
\renewcommand{\theassumption}{S\arabic{assumption}}
\renewcommand{\theremark}{S\arabic{remark}}
\renewcommand{\theHsection}{supp.\arabic{section}}
\renewcommand{\theHfigure}{supp.\arabic{figure}}
\renewcommand{\theHtable}{supp.\arabic{table}}
\renewcommand{\theHremark}{supp.\arabic{remark}}

\section{Assumptions}

We need the following regularity conditions for our asymptotic theory.

\begin{assumption}\label{a:1} $\,$
$X$ is continuously distributed on a compact
        and convex support $\mathcal{X} \subset \mathbb{R}^k$.
        The density of $X$ is bounded and bounded away from zero on $\mathcal{X}$.
\end{assumption}

\begin{assumption}\label{a23} $\,$
    \begin{enumerate}[ref =\ref{a23}.\arabic*]
        \item \label{a:2} \begin{enumerate}[label=(\roman*),ref=\ref{a:2}.(\roman*)]
            \item \label{a:2i} $D$ is independent of $\{Y(0),Y(1)\}$ conditional on $X = x$ for almost every $x$.
            \item\label{a:2ii} There exists $c>0$ such that $c<\pr(D = 1\mid X = x)<1-c$ for almost every $x$.
        \end{enumerate}
        \item\label{a:3} $\{Y_i, D_i, X_i\}^N_{i=1}$ is an independent
        and identically distributed sample of $(Y, D, X)$.
    \end{enumerate}
\end{assumption}

\begin{assumption}\label{a2p3p} $\,$
    \begin{enumerate}[ref =\ref{a2p3p}.\arabic*]
        \item \label{a:2p} \begin{enumerate}[label=(\roman*),ref=\ref{a:2p}.(\roman*)]
            \item\label{a:2pi} $D$ is independent of $Y(0)$ conditional on $X = x$ for almost every $x$.
            \item\label{a:2pii} There exists $c>0$ such that $\pr(D = 1\mid X = x) < 1 - c$ for almost every $x$.
        \end{enumerate}
        \item \label{a:3p} Conditional on $D_i = d$, $\{(Y_i, X_i)\}^{N_{d}}_{i=1}$ is an independent and identically distributed sample from the distribution of $(Y,X\mid D = d)$ for each $d\in\{0,1\}$.
        For some $r \geq 1$, $N^r_1 /N_0 \rightarrow \theta$ with $\theta\in\mathbb{R}^+$.
    \end{enumerate}
\end{assumption}

\begin{assumption}\label{a:4} $\,$
For each $d \in \{0, 1\}$,
    \begin{enumerate}[ref=\ref{a:4}.\arabic*]
            \item\label{a:4i} $x\mapsto\mu(d, x)$ and $x\mapsto\sigma^2(d, x)$
                are Lipschitz continuous in $\mathcal{X}$,
            \item\label{a:4ii} $x\mapsto\E(Y^4\mid D =d, X = x)$
                is bounded uniformly on $\mathcal{X}$, and
            \item\label{a:4iii} $x\mapsto\sigma^2(d, x)$
                is bounded away from zero on $\mathcal{X}$.
        \end{enumerate}
\end{assumption}

As is prevalent in this literature, analyses are based on Assumption \ref{a:2} that (i) assignment to treatment is independent of potential outcomes conditional on observed pretreatment variables, and (ii) there is sufficient overlap in the support of the conditional distribution of $X_i$ given $D_i = 0$ and that of the conditional distribution of $X_i$ given $D_i = 1$.
The combination of the unconfoundedness assumption in \ref{a:2i} and the overlap assumption in \ref{a:2ii} is referred to as strong ignorability \citep{rosenbaum1983central}. These two assumptions are strong and may not be satisfied in some cases. However, according to \cite{imbens2009recent}, there is no general approach to estimating treatment effects without unconfoundedness, and they describe several methods for assessing its plausibility.
The assumption of compactness and convexity of the support of the covariates in Assumption \ref{a:1} are convenient regularity conditions.
Assumption \ref{a:3} assumes that the sampling is random.
And Assumption \ref{a:4} requires weak smoothness restrictions on the conditional distribution of $Y$ given $X$.

Let $\lambda=\left(\lambda_1, \ldots, \lambda_k\right)^{\T}$ be a $k$-dimensional vector of non-negative integers.
The following assumptions are needed for deriving asymptotic properties of
$\hat{\tau}_{\bc}$ and $\hat{\tau}^t_{\bc}$
\begin{assumption}\label{a2}
$\;$
    Let $|\lambda| = \sum^k_{\ell=1} \lambda_\ell$ and $x^\lambda = \prod_{\ell=1}^k x_\ell^{\lambda_\ell} $, where $x_\ell$ is the $\ell$th element of $x$.
    Define a series $\{\lambda(q)\}^\infty_{q=1}$ that contains all distinct vectors such that $|\lambda(q)|$ is non-decreasing.
    Let $Q$ be the series length and $p^Q(x) = (x^{\lambda(1)},\ldots, x^{\lambda(Q)})^\T$.
    And let $N\mapsto Q(N)$ be an increasing function of $N$.
    A non-parametric series regression estimator
    $\hat{\mu}(d, x)$ is given by
\[
    \hat{\mu}(d, x) = p^{Q(N)}(x)^\T\left(
    \sum_{i:D_i=d} p^{Q(N)}(X_i)p^{Q(N)}(X_i)^\T \right)^- \sum_{i:D_i=d} p^{Q(N)}(X_i)Y_i,
\]
where $(\cdot)^-$ is a generalized inverse.
\end{assumption}

\begin{assumption}\label{as:3}
$\;$
\begin{enumerate}
    \item The support of $X$, $\mathcal{X}\subset \mathbb{R}^k$, is a Cartesian product of compact intervals.
    \item $Q(N)=N^\nu$, with $0<\nu < \min \{2/(4k + 3),2/(4k^2 - k)\}$.
    \item There is a $C$ such that for each multi-index $\lambda$ the $\lambda$th partial derivative of $\mu(d,x)$ exists for $d = 0, 1$ and is bounded by $C^{|\lambda|}$.
\end{enumerate}
\end{assumption}

Here, Assumption \ref{a2} summarises the conditions for the series estimator $\hat{\mu}(d, x)$ employed by \cite{abadie2011bias},
and Assumption \ref{as:3} is used to establish the asymptotic behaviour of bias-corrected rematching estimators $\hat{\tau}_{\bc}$ and $\hat{\tau}^t_{\bc}$.

\section{Proof}

Proposition \ref{prop:1} follows directly from Equation (\ref{eqt:KM}), and its proof is therefore omitted.

\subsection{Proof of Proposition \ref{prop:2}}
First, it can be proven that
\begin{align*}
    \hat{\tau} &= \frac{1}{N}\sum^N_{i=1}{\Tilde{Y}_i(1) -\Tilde{Y}_i(0)}\\
    &= \frac{1}{N}\left\{\sum_{\substack{1\le i\le N \\ D_i=1}}\left(Y_i- \frac{1}{M(i)}\sum_{j\in \mathcal{J}(i)} Y_j\right) + \sum_{\substack{1\le i\le N \\ D_i=0}}\left( \frac{1}{M(i)}\sum_{j\in \mathcal{J}(i)} Y_j -Y_i\right)\right\}
    \\
    &= \frac{1}{N}\sum_{i=1}^N(2D_i-1)Y_i-
    \sum_{\substack{1\le i\le N \\ D_i=1}}\sum_{\substack{1\le j\le N \\ D_j=0}}\frac{1}{M_0+M_{\re}(i)}\mathbb{1}\{j\in \mathcal{J}(i)\}Y_j\\&\quad+
    \sum_{\substack{1\le i\le N \\ D_i=0}}\sum_{\substack{1\le j\le N \\ D_j=1}}\frac{1}{M_0+M_{\re}(i)}\mathbb{1}\{j\in \mathcal{J}(i)\}Y_j.
\end{align*}
Note that
{\footnotesize
\begin{align*}
    \sum_{\substack{1\le i\le N \\ D_i=0}}\sum_{\substack{1\le j\le N \\ D_j=1}}&\frac{1}{M_0+M_{\re}(i)}\mathbb{1}\{j\in \mathcal{J}(i)\}Y_j\\
    =& \sum_{\substack{1\le i\le N \\ D_i=0}}\sum_{\substack{1\le j\le N \\ D_j=1}}\frac{1}{M_0+M_{\re}(i)}[\mathbb{1}\{j\in \mathcal{J}_{M_0}(i)\}+\mathbb{1}\{j\in \mathcal{L}_{\re}(i)\}]Y_j\\
    =& \sum_{\substack{1\le j\le N \\ D_j=1}}\left[\sum_{\substack{1\le i\le N \\ D_i=0}}\frac{\mathbb{1}\{j\in \mathcal{J}_{M_0}(i)\}}{M_0+M_{\re}(i)}+\sum_{\substack{1\le i\le N \\ D_i=0}}\frac{\mathbb{1}\{j\in \mathcal{L}_{\re}(i)\}}{M_0+M_{\re}(i)}\right]Y_j\\
    =& \sum_{\substack{1\le j\le N \\ D_j=1}}\left[\sum_{\substack{1\le i\le N \\ D_i=0}}\frac{\mathbb{1}\{j\in \mathcal{J}_{M_0}(i)\}}{M_0+M_{\re}(i)}+\sum_{\substack{1\le i\le N \\ D_i=0}}\frac{\mathbb{1}\{i\in \mathcal{J}_{\re}(j)\}\mathbb{1}\{K_{M_0}(j)=0\}}{M_0+M_{\re}(\mathcal{J}_{\re}(j))}\right]Y_j\\
    =& \sum_{\substack{1\le j\le N \\ D_j=1}}\left[\sum_{\substack{1\le i\le N \\ D_i=0}}\frac{\mathbb{1}\{j\in \mathcal{J}_{M_0}(i)\}}{M_0+M_{\re}(i)}+\frac{\mathbb{1}\{K_{M_0}(j)=0\}}{M_0+M_{\re}(\mathcal{J}_{\re}(j))}\right]Y_j\\
    =& \sum_{\substack{1\le j\le N \\ D_j=1}}\frac{K(j)}{M_0}Y_j,
\end{align*}}
and similarly, we have
\[
    \sum_{\substack{1\le i\le N \\ D_i=1}}\sum_{\substack{1\le j\le N \\ D_j=0}}\frac{1}{M_0+M_{\re}(i)}\mathbb{1}\{j\in \mathcal{J}(i)\}Y_j = \sum_{\substack{1\le j\le N \\ D_j=0}}\frac{K(j)}{M_0}Y_j.
\]
Hence,
\begin{align*}
     \hat{\tau} &= \frac{1}{N}\sum_{i=1}^N(2D_i-1)Y_i-
    \sum_{\substack{1\le j\le N \\ D_j=0}}\frac{K(j)}{M_0}Y_j+
    \sum_{\substack{1\le j\le N \\ D_j=1}}\frac{K(j)}{M_0}Y_j
    \\
    &= \frac{1}{N}\sum_{i=1}^N(2D_i-1)\left\{1+\frac{K(i)}{M_0}\right\}Y_i  .
\end{align*}
Similarly, we can prove that
\[\hat{\tau}^t
    =  \frac{1}{N_1}\sum_{i=1}^N \left\{D_i - (1-D_i)
     \frac{K(i)}{M_0}\right\}Y_i,\]
which is the desired result.\qed

\subsection{Proof of Theorem \ref{thm:1}}\label{Sec:Thm1Proof}
The proof below follows a similar argument as in the proof of Theorem 1(i) in \cite{abadie2006large}.
Recall from Section \ref{sec:bias}
that the bias term is given by
\begin{align*}
B &=   \frac{1}{N}\sum^N_{i=1}(2D_i - 1)\left[ \frac{1}{M(i) }\sum_{j\in \mathcal{J}(i)} \left\{\mu(1-D_i,X_i) - \mu\left(1-D_i,X_{j}\right)\right\}\right]\\
&= \frac{1}{N}\sum_{i=1}^N \sum_{j\in \mathcal{J}(i)} \frac{B_{j,i}}{M(i)},
\end{align*}
where $M(i)=|\mathcal{J}(i)|$, and
\[ B_{j,i}= (2D_i - 1) \left\{\mu(1-D_i,X_i) - \mu\left(1-D_i,X_{j}\right)\right\}
\]
is the unit-level bias for the match with index $j$ of the unit $i$.
Let $U_{j ,i}=X_j-X_i$ be the unit-level matching discrepancy.
Then $B_{j,i}$ can be rewritten as follows:
{\small
\begin{align*}
    B_{j ,i} &= D_i\{\mu(0,X_i)-\mu(0,X_{j})\}-(1-D_i)\{\mu(1,X_i)-\mu(1,X_{j})\}\\
    &=D_i\{\mu(0,X_i)-\mu(0,X_i+U_{j ,i})\}-(1-D_i)\{\mu(1,X_i)-\mu(1,X_i+U_{j ,i})\},
\end{align*}}where $j\in \mathcal{J}(i)$.
Next, we prove that $M(i)$ is bounded in probability.
Recall that $M(i)= M_0+  M_{\re}(i),$ where $ M_{\re}(i)=\sum^N_{j=1}\mathbb{1}\{i \in \mathcal{J}_{\re}(j)\}$.
By the definition of $\mathcal{J}_{\re}(i)$, we know that $\mathcal{J}_{\re}(j) \subseteq \mathcal{J}_{1}(j)$ for all $j=1,\ldots,N$.
So, we have $\mathbb{1}\{i \in \mathcal{J}_{\re}(j)\} \leq \mathbb{1}\{i \in \mathcal{J}_{1}(j)\}$ for all $i,j=1,\ldots,N,$ implying that
\[ M_{\re}(i) \leq \sum_{j=1}^N\mathbb{1}\{i \in \mathcal{J}_{1}(j)\} \equiv K_1(i),
\]
where the inequality essentially means that the number of times unit $i$ is used as the nearest match among \emph{unmatched}  units must be bounded above by the number of times unit $i$ is used as the nearest match among \emph{all} units.
According to Lemma~3(i) of \cite{abadie2006large}, we have $K_1(i)=O_p(1)$.
Hence, $M(i)\leq M_0+K_{1}(i) =O_p(1)$.

Let $J(i)=\argmax_{j\in \mathcal{J}(i)} B_{j ,i}$, $i=1,\ldots,N$, denoting the index of the farthest match in $\mathcal{J}(i)$.
We obtain
\begin{align*}
E(N^{2 / k}B^2)
= & N^{2 / k}E\left(\frac{1}{N}
        \sum_{i=1}^{N}\sum_{j\in \mathcal{J}(i)}\frac{B_{j ,i}}{M(i)}\right)^2\\
\leq & N^{2 / k}E\left(\frac{1}{N }
        \sum_{i=1}^{N}\sum_{j\in \mathcal{J}(i)}\frac{|B_{J(i) ,i}|}{M(i)}\right)^2 \\
    = & N^{2 / k}E\left(\frac{1}{N }
        \sum_{i=1}^{N}  |B_{J(i) ,i}| \right)^2.
\end{align*}

Under the Lipschitz continuity assumption on $\mu(d,X)$ for $d \in\{0,1\}$, we have $|B_{j ,i}| \leq C_1\lVert U_{j ,i}\rVert$,
where $C_1$ is some positive constant that is finite by assumption.
Hence, we get
\begin{align}
E(N^{2 / k}B^2)
\leq & N^{2 / k} E\left(\frac{1}{N} \sum_{i=1}^N  C_1\|U_{J(i), i}\|\right)^2  \nonumber \\
= & N^{2 / k}C_1^2 \frac{1}{N^2} \sum_{i=1}^N  \sum_{j=1}^N E(\|U_{J(i), i}\| \|U_{J(j), j}\|) \nonumber \\
\leq & N^{2 / k}C_1^2 \frac{1}{N^2} \sum_{i=1}^N  \sum_{j=1}^N E\left(\frac{\|U_{J(i), i}\|^2+ \|U_{J(j), j}\|^2}{2}\right)\nonumber \\
= & N^{2 / k}C_1^2 \frac{2}{N^2}  E\left(\sum_{i=1}^N  \frac{N}{2}\|U_{J(i), i}\|^2\right)\nonumber \\
= & N^{2 / k-1}C_1^2 \sum_{i=1}^N E \|U_{J(i), i}\|^2\nonumber \\
= &  N^{2 / k-1} C_1^2 E\left\{\frac{1}{N_0^{2 / k}} \sum_{i:D_i=1} E(N_0^{2 / k}\left\|U_{J(i), i}\right\|^2 \mid D_1, \ldots, D_N, X_i)\right. \nonumber \\
& \left.+\frac{1}{N_1^{2 / k}} \sum_{i:D_i=0} E(N_1^{2 / k}\left\|U_{J(i), i}\right\|^2 \mid D_1, \ldots, D_N, X_i)\right\}, \label{eqt:expectation_of_B_bound}
\end{align}
where the last equality holds using the tower property.

According to Lemma~2 of \cite{abadie2006large}, if Assumption \ref{a:1} holds, then the normalized moments of the matching discrepancies, i.e., $N^{1/k} \|U_{j, i}\|$ for $j\in\mathcal{J}_m(i)$ and $m<\infty$, are bounded for all points in the support.
However, for our proposed rematching estimator,
the matched index set for the $i$th unit,
i.e., $\mathcal{J}(i)$, is not equal to any
$\mathcal{J}_m(i)$ with a fixed $m$.
To derive an upper bound for $\| U_{J(i),i} \|\leq \sup_{j\in\mathcal{J}(i)}\|X_i-X_j\|$, we will bound the matching discrepancies and rematching discrepancies separately.
By definition,
$\mathcal{J}(i)=\mathcal{J}_{M_0}(i)\cup\mathcal{L}_{\re}(i)$.
So, it holds that
\begin{align}\label{eqt:decompose_supDistPJi}
\sup_{j \in \mathcal{J}(i)}\|X_{i}-X_{j}\|\leq
\sup_{j \in \mathcal{J}_{M_0}(i)}\|X_{i}-X_{j}\|+ \sup_{j \in \mathcal{L}_{\re}(i)}\|X_{i}-X_{j}\|,
\end{align}
where $\sup_{j\in\emptyset}\|X_{i}-X_{j}\| := 0$.
Since $\mathcal{L}_{\re}(i) = \{\ell \in  \{1,\ldots, N\} : i \in \mathcal{J}_{\re}(\ell)
\}$, we know that
$j \in \mathcal{L}_{\re}(i)$ implies
$i \in \mathcal{J}_{\re}(j)$.
Also,  note that
\begin{align*}
    \mathcal{J}_{\re}(j)
        &= \left\{
        \begin{array}{cc}
             \mathcal{J}_1(j) &  \text{if $K_{M_0}(j) = 0$}; \\
             \emptyset & \text{if $K_{M_0}(j) > 0$},
        \end{array}
        \right. \\
        & \subseteq \mathcal{J}_1(j).
\end{align*}
Hence, $j \in \mathcal{L}_{\re}(i)$ further implies $i\in \mathcal{J}_1(j)$.
Then,
let $\mathcal{S}(i) = \{ j \;:\; i\in\mathcal{J}_1(j)\}$.
We can bound (\ref{eqt:decompose_supDistPJi}) as follows:
\begin{align}\label{eqt:bound_supXX}
	\sup_{j \in \mathcal{J}(i)}\|X_{i}-X_{j}\|
	&\leq \sup_{j \in \mathcal{J}_{M_0}(i)}\|X_{i}-X_{j}\|
			+ \sup_{j \in \mathcal{S}(i) } \|X_{i}-X_{j}\| .
\end{align}
Now let $\kappa(k)$ be the kissing number in dimension $k$, i.e.,
the maximum number of non-overlapping unit spheres that can touch a common unit sphere.
By considering the ratio of volumes of $k$-dimensional spheres with radii 1 and 3,
it is not hard to see that $\kappa(k) \leq 3^k$ is bounded for any finite $k$; see, e.g., \cite{CohnZhao2014} for a sharper bound.
Note that a particular point $X_i$ can only accommodate a limited number of the points
regarding $X_i$ as their nearest neighbor due to spatial constraints.
The value of $|\mathcal{S}(i)|$ is maximized when
we arrange the maximum number of unit spheres around a unit sphere centered at $X_i$
in a way that each sphere touches the central sphere without overlapping.
By definition, this maximum number is the kissing number; see, e.g., \cite{ivaRonaldSeema2006} and \cite{zong1998} for more details.
In other words, the set $\mathcal{S}(i)$ contains at most $\kappa(k)$ points, i.e., $|\mathcal{S}(i)| \leq \kappa(k)\leq 3^k$.

Hence, $j \in \mathcal{S}(i)$ implies $j \in \mathcal{J}_{3^k}(i)$.
Consequently, (\ref{eqt:bound_supXX}) can be further bounded as follows:
\begin{align}\label{eqt:bound_supXX2}
	\sup_{j \in \mathcal{J}(i)}\|X_{i}-X_{j}\|
	&\leq \sup_{j \in \mathcal{J}_{M_0}(i)}\|X_{i}-X_{j}\|
			+ \sup_{j \in \mathcal{J}_{3^k}(i) } \|X_{i}-X_{j}\| .
\end{align}
Each term on the right-hand side of (\ref{eqt:bound_supXX2}) is bounded
in view of Lemma~2 of \cite{abadie2006large}.
Moreover, the conditional expectations $E(N_0^{2 / k}\left\|U_{J(i), i}\right\|^2 \mid D_1, \ldots, D_N, X_i)$ and $E(N_1^{2 / k}\left\|U_{J(i), i}\right\|^2 \mid D_1, \ldots, D_N, X_i)$
are both bounded by a constant $C$ that does not depend on $i$ according to Lemma~2 of \cite{abadie2006large}.
Thus, using this uniform bound, we have
\begin{align*}
    \sum_{i:D_i=1} E(N_0^{2 / k}\left\|U_{J(i), i}\right\|^2 \mid D_1, \ldots, D_N, X_i) &\leq C N_1 , \\
    \sum_{i:D_i=0} E(N_1^{2 / k}\left\|U_{J(i), i}\right\|^2 \mid D_1, \ldots, D_N, X_i) &\leq C N_0.
\end{align*}
Therefore, (\ref{eqt:expectation_of_B_bound}) can be bounded as follows:
\begin{align*}
E(N^{2 / k}B^2)
\leq & C_2 E\left\{\left(\frac{N}{N_0}\right)^{2 / k} \frac{N_1}{N}+\left(\frac{N}{N_1}\right)^{2 / k} \frac{N_0}{N}\right\}
\end{align*}
for some positive constant $C_2$. By Chernoff's inequality, any moment of $N / N_1$ or $N / N_0$ is uniformly bounded in $N$, as in \cite{abadie2006large}.
Part 1 of the theorem follows from Markov's inequality.

The proof of part 2 is similar to that of part 1 and is omitted.\qed

\subsection{Proof of Lemma \ref{lem:1}}
By Lemma~3(i) of \citet{abadie2006large}, $K_{M_0}(i) = \sum^N_{\ell=1}\mathbb{1}\{i \in \mathcal{J}_{M_0}(\ell)\} = O_p(1)$.
Consider
{\small
\begin{align*}
    K(i) &= \left[\sum^N_{\ell=1}\mathbb{1}\{i \in \mathcal{J}_{M_0}(\ell)\} \frac{M_0}{M_{\re}(\ell)+M_0}\right]
    +\mathbb{1}\{K_{M_0}(i)=0\} \frac{M_0}{M_{\re}(\mathcal{J}_{\re}(i))+M_0}\\
    &=
\begin{cases}
  \sum^N_{\ell=1}\mathbb{1}\{i \in \mathcal{J}_{M_0}(\ell)\} {M_0}/\{M_{\re}(\ell)+M_0\} & \text{if }K_{M_0}(i)\neq 0;
    \\
    {M_0}/\{M_{\re}(\mathcal{J}_{\re}(i))+M_0\} & \text{if }K_{M_0}(i)= 0,
    \\
\end{cases}
\end{align*}}
where ${M_0}/\{M_{\re}(i)+M_0\}\in(0,1]$
is the adjusted weight for unit $i$ after rematching and ${M_0}/\{M_{\re}(\mathcal{J}_{\re}(i))+M_0\}$ is the weight given, after rematching, to unit $i$ that is unmatched in the simple matching.
When $K_{M_0}(i)\neq 0$, we have
\begin{equation}
    K(i)=
    \sum^N_{\ell=1}\mathbb{1}\{i \in \mathcal{J}_{M_0}(\ell)\}  \frac{M_0}{M_{\re}(\ell)+M_0}\leq K_{M_0}(i) = O_p(1).\nonumber
\end{equation} And when $K_{M_0}(i)= 0$,
\begin{equation}
    K(i)=\frac{M_0}{M_{\re}(\mathcal{J}_{\re}(i))+M_0}
    \in(0,1)
    . \nonumber
\end{equation}
Therefore, $K(i)=O_p(1)$.

According to Lemma~3 of \citet{abadie2006large}, the $q$th moment of $K_{M_0}(i)$ conditional on $X_i = x, D_{1:N}$, which is  $\E\{K_{M_0}(i)^q\mid X_i = x, D_{1:N}\}$, is uniformly bounded in $N$.
Since $K(i)$ is either $\leq K_{M_0}(i)$ or $\in(0,1)$,
$\E\{K(i)^{q}\mid X_i = x, D_{1:N}\}$ will also be bounded in $N$, so the first part of the lemma follows.

Part 2 of Lemma \ref{lem:1} can be proven using the same argument as for $E\{K(i)^q\}$.

For part 3 of Lemma \ref{lem:1}, because the moments of $K(i)$ are bounded uniformly in $N$, and
because the variance $\sigma^2(d,x)$
is bounded by $\bar{\sigma}^2=\sup_{x,d}\sigma^2(d,x)$, which is finite by Assumption \ref{a:4}, $\E[\{1+K(i)/M_0\}^2\sigma^2(d,x)]$ is bounded by $\bar{\sigma}^2E[\{1+K(i)/M_0\}^2]$ and is finite. Hence $E(V^R) = O(1)$.\qed

\subsection{Proof of Theorem \ref{thm:generalVar}}

Before proving Theorem \ref{thm:generalVar}, we need some preliminary results given by \cite{abadie2002simple}.
First, let $\bar{r}_{d}(x)$ be the number of units with $D_{i}=d$ and $X_{i} \geq x$.
Define $X_{(i, k)}=X_{j}$ if $\bar{r}_{D_{i}}(X_{i})-\bar{r}_{D_{i}}(X_{j})=k$, and $\bar{r}_{D_{i}}(X_{i})-\lim _{x \uparrow X_{j}} \bar{r}_{D_{i}}(x)=k-1$.
When $X$ is a scalar,
$A_{M_0}(i)=(X_{i} / 2+X_{(i,-M_0)} / 2, X_{i} / 2+X_{(i, M_0)} / 2).$
Then, the exact conditional distribution of $K_{M_0}(i)$ is,
\[
K_{M_0}(i) \mid \mathbf{D},\{X_{j}\}_{D_{j}=1}, D_{i}=1 \sim \Bin\left(N_{0}, \int_{A_{M_0}(i)} f_{0}(z) d z\right),
\]
\[
K_{M_0}(i) \mid \mathbf{D},\{X_{j}\}_{D_{j}=0}, D_{i}=0 \sim \Bin\left(N_{1}, \int_{A_{M_0}(i)} f_{1}(z) d z\right).
\]
Next, let \textnormal{Gamma}($\alpha$, $\beta$) denote the gamma distribution with a shape parameter $\alpha$ and a scale parameter $\beta$.
Given $X_{i}=x$, and $D_{i}=1$,
\[
2 N_{1} \cdot \frac{f_{1}(x)}{f_{0}(x)} \cdot \int_{A_{M_0}(i)} f_{0}(z) d z \rightarrow \textnormal{Gamma}(2 M_0, 1)
\] in distribution,
and given $X_{i}=x$ and $D_{i}=0$,
\[
2 N_{0} \cdot \frac{f_{0}(x)}{f_{1}(x)} \cdot \int_{A_{M_0}(i)} f_{1}(z) d z \rightarrow \textnormal{Gamma}(2 M_0, 1)
\] in distribution.

Note that
\begin{align*}
N\var(\hat{\tau} )
=& \E(V^R) + V^{\tau(X)}\\
=& \E\left[\left\{1+\frac{K(i)}{M_0}\right\}^2\sigma^2(D_i,X_i)\right]+V^{\tau(X)}\\
=&
\E\left[\left\{1+\frac{K(i)}{M_0}\right\}^2\sigma^2(1,X_i)\mid D_i=1\right]p \\
&+ \E\left[\left\{1+\frac{K(i)}{M_0}\right\}^2\sigma^2(0,X_i)\mid D_i=0\right](1-p)+V^{\tau(X)},
\end{align*}
where $p$ is the probability of treatment assignment.
Similarly,
\begin{align*}
N\var(\hat{\tau}_{0} ) =& \E(V^{R}_0) + V^{\tau(X)}\\
=&  \E\left[\left\{1+\frac{K_{M_0}(i)}{M_0}\right\}^2\sigma^2(1,X_i)\mid D_i=1\right]p\\
&+ \E\left[\left\{1+\frac{K_{M_0}(i)}{M_0}\right\}^2\sigma^2(0,X_i)\mid D_i=0\right](1-p)+V^{\tau(X)}.
\end{align*}
The number of times a unit is used as a match is calculated as
\[
    K_{M_0}(i) = \sum^N_{\ell=1}\mathbb{1}\{i \in \mathcal{J}_{M_0}(\ell)\}
\]
when simple matching estimators are used.
Recall that $M(\ell) = M_0 + M_{\re}(\ell)$.
By Proposition 1,
\[
K(i) = \sum^N_{\ell=1}\mathbb{1}\{i \in \mathcal{J}(\ell)\}  \frac{M_0}{M(\ell)},
\]
which is not comparable with $K_{M_0}(i)$ as it is a weighted form.
To obtain a comparable form, we let
\[
H(i) = \sum^N_{\ell=1}\mathbb{1}\{i \in \mathcal{J}(\ell)\} .
\]
It is the number of times a unit is used as a match when rematching estimators are used.
Equivalently,
\[
H(i) = \max\{K_{M_0}(i),1\},
\]
since if a unit remains unmatched after the first matching, it will be matched to its nearest neighbour in the rematching process.

The conditional expectation of $H(i)$ is
\begin{align*}
&\E[H(i) \mid \mathbf{D},\{X_{j}\}_{D_{j}=1}, D_{i}=1]\\
&= \E[ \max\{K_{M_0}(i),1\} \mid \mathbf{D},\{X_{j}\}_{D_{j}=1}, D_{i}=1]\\
&=
E[K_{M_0}(i)\mathbb{1}\{K_{M_0}(i)\geq1\} \mid \mathbf{D},\{X_{j}\}_{D_{j}=1}, D_{i}=1] \\
&\qquad\qquad+ E[\mathbb{1}\{K_{M_0}(i)<1\} \mid \mathbf{D},\{X_{j}\}_{D_{j}=1}, D_{i}=1]\\
&= E[K_{M_0}(i)\mid \mathbf{D},\{X_{j}\}_{D_{j}=1}, D_{i}=1] \\
&\qquad\qquad+ \pr[K_{M_0}(i)<1 \mid \mathbf{D},\{X_{j}\}_{D_{j}=1}, D_{i}=1],
\end{align*}
and the conditional expectation of $H^2(i)$  is
\begin{align*}
&\E[H^2(i) \mid \mathbf{D},\{X_{j}\}_{D_{j}=1}, D_{i}=1]\\
&= E[K_{M_0}^2(i)\mid \mathbf{D},\{X_{j}\}_{D_{j}=1}, D_{i}=1] \\
&\qquad\qquad+ \pr[K_{M_0}(i)<1 \mid \mathbf{D},\{X_{j}\}_{D_{j}=1}, D_{i}=1]
\end{align*}

Define $P_{M_0}(i)=D_{i} \cdot \int_{A_{M_0}(i)} f_{0}(z) d z+(1-D_{i}) \cdot \int_{A_{M_0}(i)} f_{1}(z) d z$. According to \cite{abadie2002simple},
\[
\E[K_{M_0}(i) \mid \mathbf{D},\{X_{j}\}_{D_{j}=1}, D_{i}=1]=N_{0} P_{M_0}(i),
\]
\[
\E[K_{M_0}^{2}(i) \mid \mathbf{D},\{X_{j}\}_{D_{j}=1}, D_{i}=1]=N_{0} P_{M_0}(i)+N_{0}(N_{0}-1) P_{M_0}(i)^{2},
\]
and therefore,
\begin{align*}
\E& \left.\left[\left\{1+\frac{K_{M_0}(i)}{M_0}\right\}^{2}\sigma^{2}(1,X_{i}) \right\rvert D_{i}=1\right] \\
& =\E\left(\left[1+\frac{1}{M^{2}}\{N_{0} P_{M_0}(i)+N_{0}(N_{0}-1) P_{M_0}(i)^{2}\}\right.\right.\\
&\qquad\qquad+\left.\left.\left.\frac{2}{M_0} N_{0} P_{M_0}(i)\right] \sigma^{2}(1,X_{i}) \right\rvert\, D_{i}=1\right)\\
&=\E\left(\left[1+\frac{1}{M_0}\left\{\frac{(1-p)}{p} \frac{f_{0}(X_{i})}{f_{1}(X_{i})}+\frac{(1-p)^{2}}{2 p^{2}} \frac{f_{0}(X_{i})^{2}}{f_{1}(X_{i})^{2}}(2 M+1)\right\}\right.\right.\\
&\qquad\qquad+\left.\left.\left.\frac{2(1-p)}{p} \frac{f_{0}(X_{i})}{f_{1}(X_{i})}\right] \sigma^{2}(1,X_{i}) \right\rvert\, D_{i}=1\right)+o(1).
\end{align*}

Similarly, we derive that
\begin{align*}
\E[H(i) &\mid \mathbf{D},\{X_{j}\}_{D_{j}=1}, D_{i}=1]=N_{0} P_{M_0}(i)+ \{1- P_{M_0}(i)\}^{N_{0}},\\
\E[H^2(i) &\mid \mathbf{D},\{X_{j}\}_{D_{j}=1},
D_{i}=1]\\
&= N_{0} P_{M_0}(i)
+N_{0}(N_{0}-1) P_{M_0}(i)^{2} + \{1- P_{M_0}(i)\}^{N_{0}}.
\end{align*}
And
\begin{align*}
\E&\left.\left[\left\{1+\frac{K(i)}{M_0}\right\}^{2}\sigma^{2}(1,X_{i}) \right\rvert D_{i}=1\right] \\
&\leq \left.\E\left[\left\{1+\frac{H(i)}{M_0}\right\}^{2}\sigma^{2}(1,X_{i}) \right\rvert D_{i}=1\right] \\
& =\E\left\{\left(1+\frac{1}{M^{2}}[N_{0} P_{M_0}(i)+N_{0}(N_{0}-1) P_{M_0}(i)^{2} + \{1- P_{M_0}(i)\}^{N_{0}}]\right.\right.\\
&\qquad\left.\left.\left.
+\frac{2}{M_0} \{N_{0} P_{M_0}(i) + (1- P_{M_0}(i))^{N_{0}} \}\right) \sigma^{2}(1,X_{i}) \right\rvert D_{i}=1\right\} .
\end{align*}
The first inequality holds because $K(i)/M_0=\sum^N_{\ell=1}\mathbb{1}\{i \in \mathcal{J}(\ell)\}  /M(\ell)$, $H(i)/M_0=\sum^N_{\ell=1}\mathbb{1}\{i \in \mathcal{J}(\ell)\}  /M_0$, and $M(\ell)\geq M_0$ for all $\ell$.
Since $\{1- P_{M_0}(i)\}\in[0,1)$ when $M<\min(N_0,N_1)$, we know that $\{1- P_{M_0}(i)\}^{N_{0}} = o(1)$.
So, similarly, we know that
\begin{align*}
\E&\left[\left(1+\frac{H(i)}{M_0}\right)^{2}\sigma^{2}(1,X_{i}) \mid D_{i}=1\right]\\
&=\E\left[\left.\left(1+\frac{1}{M_0}\left\{\frac{(1-p)}{p} \frac{f_{0}(X_{i})}{f_{1}(X_{i})}+\frac{(1-p)^{2}}{2 p^{2}} \frac{f_{0}(X_{i})^{2}}{f_{1}(X_{i})^{2}}(2 M+1)\right\}\right.\right.\right.\\
&\qquad\left.\left.\left.+\frac{2(1-p)}{p} \frac{f_{0}(X_{i})}{f_{1}(X_{i})}\right) \sigma^{2}(1,X_{i}) \right\rvert\, D_{i}=1\right]+o(1)\\
&=
\E \left.\left[\left\{1+\frac{K_{M_0}(i)}{M_0}\right\}^{2}\sigma^{2}(1,X_{i}) \right\rvert D_{i}=1\right].
\end{align*}
That is,
\begin{align*}
\E&\left.\left[\left\{1+\frac{K(i)}{M_0}\right\}^{2}\sigma^{2}(1,X_{i}) \right\rvert D_{i}=1\right]\\
&\qquad\leq
\E \left.\left[\left\{1+\frac{K_{M_0}(i)}{M_0}\right\}^{2}\sigma^{2}(1,X_{i}) \right\rvert D_{i}=1\right].
\end{align*}
Similarly,
\begin{align*}
\E&\left[\left\{1+\frac{K(i)}{M_0}\right\}^2\sigma^2(0,X_i)\mid D_i=0\right]\\
&\qquad\leq\E\left[\left\{1+\frac{K_{M_0}(i)}{M_0}\right\}^2\sigma^2(0,X_i)\mid D_i=0\right].
\end{align*}
Then, it is obvious that $N\var(\hat{\tau}) \leq N\var(\hat{\tau}_{0}) $. \qed

\subsection{Proof of Theorem \ref{thm:4}}
Note that
\begin{align*}
&N\var(\hat{\tau} )   \\
&= \E(V^R) + V^{\tau(X)}\\
&= \E\left[\left\{1+\frac{K(i)}{M_0}\right\}^2\sigma^2(D_i,X_i)\right]+V^{\tau(X)}\\
&=
\E\left[\left\{1+\frac{K(i)}{M_0}\right\}^2\sigma^2(1,X_i)\mid D_i=1\right]p \\
&\quad+ \E\left[\left\{1+\frac{K(i)}{M_0}\right\}^2\sigma^2(0,X_i)\mid D_i=0\right](1-p)+V^{\tau(X)},
\end{align*}
where $p$ is the probability of treatment assignment.
Similarly,
\begin{align}
N\var(\hat{\tau}_{0} ) \quad=&\quad \E(V^{R}_0) + V^{\tau(X)}\nonumber\\
\quad=
& \quad \E\left[\left\{1+\frac{K_{M_0}(i)}{M_0}\right\}^2\sigma^2(1,X_i)\mid D_i=1\right]p\nonumber\\
&\quad+ \E\left[\left\{1+\frac{K_{M_0}(i)}{M_0}\right\}^2\sigma^2(0,X_i)\mid D_i=0\right](1-p)+V^{\tau(X)}\nonumber.
\end{align}
It is known that
\[K_{M_0}(i)= \sum^N_{\ell=1}\mathbb{1}\{i \in \mathcal{J}_{M_0}(\ell)\} \]
and
\begin{align*}
    \frac{K(i)}{M_0}&= \sum^N_{\ell=1}\mathbb{1}\{i \in \mathcal{J}_{M_0}(\ell)\}  \frac{1}{M_{\re}(\ell)+M_0}
    +\mathbb{1}\{K_{M_0}(i)=0\}  \frac{1}{M_{\re}(\mathcal{J}_{\re}(i))+M_0}\\
    &=
\begin{cases}
  \sum^N_{\ell=1}\mathbb{1}\{i \in \mathcal{J}_{M_0}(\ell)\}  {1}/\{M_{\re}(\ell)+M_0\} & \text{if }K_{M_0}(i)\neq 0;
    \\
    {1}/\{M_{\re}(\mathcal{J}_{\re}(i))+M_0\} & \text{if }K_{M_0}(i)= 0
    .\\
\end{cases}
\end{align*}
Then we state the results below given the values of $N_0$ and $N_1$.
Under the assumptions that $k = 1$,
$f_0(x)=f_1(x)$,
and $M_0N_1<N_0$,
we have, for $D_i=1$,
\[
K(i) = K_{M_0}(i) \sim \Bin(M_0N_0,1/N_1).
\]
For $D_i=0$, given that $f_0(x)=f_1(x)$,
the term $K_{M_0}(i)$ can only take values in $\{0,1\}$ when there exist unmatched points. When $K_{M_0}(i)=1$ for the control unit $i$, exactly one of $\mathbb{1}\{i \in \mathcal{J}_{M_0}(\ell)\}$ ($\ell = 1,\ldots,N$) is $1$.
Since the number of matched control is $M_0N_1$,
we have
$K_{M_0}(i)\sim \Bern(M_0N_1/N_0),$
and
\begin{eqnarray}
    \frac{K(i)}{M_0}
    &=&
\begin{cases}
  \sum^N_{\ell=1}\mathbb{1}\{i \in \mathcal{J}_{M_0}(\ell)\}  {1}/\{M_{\re}(\ell)+M_0\} & \text{if }K_{M_0}(i) = 1;
    \\
    {1}/\{M_{\re}(\mathcal{J}_{\re}(i))+M_0\} & \text{if }K_{M_0}(i)= 0
    .\\
\end{cases}
    \nonumber
\end{eqnarray}
And also,
\begin{align*}
[K_{M_0}(i)\mid D=1]&\sim \Bin(M_0N_0,1/N_1), \\
[\mathbb{1}\{K_{M_0}(i)=1\} \mid D=0] &\sim \Bern(M_0N_1/N_0),\\
[M_{\re}(\ell)\mid K_{M_0}(i)=1, D=0]&\sim \Bin(N_0-M_0N_1,1/N_1),\\
[\{M_{\re}(\mathcal{J}_{re}(i))-1\}\mid K_{M_0}(i)=0, D=0]&\sim \Bin(N_0-M_0N_1-1,1/N_1).
\end{align*}
Denoting $[K_{M_0}(i)\mid D=1]$ as $K$, $[\mathbb{1}\{K_{M_0}(i)=1\}\mid D=0]$ as $I$, $[M_{\re}(\ell)\mid K_{M_0}(i)=1, D=0]$ and $[M_{\re}(\mathcal{J}_{re}(i))-1\mid K_{M_0}(i)=0, D=0]$ as $[H\mid I]$, we have
\begin{align*}
K&\sim \Bin(M_0N_0,1/N_1),\\
I&\sim \Bern(M_0N_1/N_0),\\
[H \mid I]&\sim \Bin(N_0-M_0N_1-\mathbb{1}(I=0),1/N_1).
\end{align*}
Now, using the assumption that $\sigma^2(d,x) = \sigma^2_d$ is a constant as a function of $x$ for each $d\in\{0,1\}$, we have that
{\footnotesize
\begin{align*}
N\var(\hat{\tau} )
&\rightarrow  \E\left[\left\{1+\frac{K(i)}{M_0}\right\}^2\sigma^2(1,X_i)\mid D_i=1\right]p \\
&\quad+ \E\left[\left\{1+\frac{K(i)}{M_0}\right\}^2\sigma^2(0,X_i)\mid D_i=0\right](1-p)+V^{\tau(X)}\\
&= \sigma^2_0(1-p)\E\left[\left\{1+
\sum^N_{\ell=1} \frac{\mathbb{1}\{i \in \mathcal{J}_{M_0}(\ell)\}}{M_{\re}(\ell)+M_0}+\frac{\mathbb{1}\{K_{M_0}(i)=0\}}{M_{\re}(\mathcal{J}_{\re}(i))+M_0}
\right\}^2\mid D_i=0\right]\\
&\quad+\sigma_1^2p\E\left[\left\{1+\frac{K(i)}{M_0}\right\}^2\mid D_i=1\right]+V^{\tau(X)} \\
&= \sigma_1^2 p\E\left(1+\frac{K}{M_0}\right)^2
+ \sigma_0^2(1-p)\E\left(1+
\frac{I}{H+M_0}
    +\frac{1-I}{H+1+M_0}
\right)^2 +V^{\tau(X)}.
\end{align*} }
For comparison, $N\var(\hat{\tau}_{0} )$ can be written as
\begin{align*}
N\var(\hat{\tau}_{0} )
&\rightarrow\E\left[\left\{1+\frac{K_{M_0}(i)}{M_0}\right\}^2\sigma^2(1,X_i)\mid D_i=1\right]p \\
&\quad+ \E\left[\left\{1+\frac{K_{M_0}(i)}{M_0}\right\}^2\sigma^2(0,X_i)\mid D_i=0\right](1-p)+V^{\tau(X)}\\
&=\sigma^2_1p\E\left(1+\frac{K}{M_0}\right)^2 + \sigma^2_0(1-p)\E\left(1+\frac{I}{M_0}\right)^2+V^{\tau(X)}.
\end{align*}
Similarly, we have
\begin{align*}
    N_1\var(\hat{\tau}^{t})
        &\rightarrow\sigma^2_1p + \sigma^2_0(1-p) \E\left(\frac{I}{H+M_0} +\frac{1-I}{H+1+M_0}\right)^2
            +V^{\tau(X),t} , \\
    N_1\var(\hat{\tau}_{0}^{t})
        &\rightarrow\sigma^2_1p + \sigma^2_0(1-p)\E\left(\frac{I}{M_0}\right)^2+V^{\tau(X),t}.
\end{align*}
Then the desired results follow.\qed

\subsection{Proof of Proposition \ref{prop:3}}
By Theorem \ref{thm:4},
we have
{\footnotesize
    \begin{align*}
        N\{\var(\hat{\tau}_{0}^{t}) - \var(\hat{\tau}^{t})\}
            &\rightarrow \sigma^2_0(1-p)\left\{\E\left(\frac{I}{M_0}\right)^2-\E\left(\frac{I}{H+M_0} +\frac{1-I}{H+1+M_0}\right)^2\right\}.
    \end{align*}}
Note that
\[
\E\left(\frac{I}{M_0}\right)^2 = \frac{1}{M_0^2} \frac{M_0 N_1}{N_0}=\frac{N_1}{N_0M_0},
\]
and
{\footnotesize
\begin{align*}
&\E\left(\frac{I}{H+M_0} +\frac{1-I}{H+1+M_0}\right)^2\\
&\quad= \E\left\{\left(\frac{1}{H+M_0}\right)^2\mid I=1 \right\} \pr(I=1)
+ \E\left\{\left(\frac{1}{H+1+M_0}\right)^2 \mid I=0 \right\} \pr(I=0)\\
&\quad=\E\left(\frac{1}{H_1+M_0}\right)^2 \frac{M_0 N_1}{N_0}
+ \E\left(\frac{1}{H_0+1+M_0}\right)^2 \left(1-\frac{M_0 N_1}{N_0}\right)\\
&\quad=\E\left(\frac{M_0}{H_1+M_0}\right)^2 \frac{N_1}{N_0M_0}
+ \E\left(\frac{M_0}{H_0+1+M_0}\right)^2 \left(\frac{1}{M_0^2}-\frac{N_1}{N_0M_0}\right).\\
\end{align*}}
We have
{\footnotesize
\begin{align*}
    &N\{\var(\hat{\tau}_{0}^{t}) - \var(\hat{\tau}^{t})\} \\
        &\quad= \sigma^2_0(1-p)\left[
        \frac{N_1}{N_0M_0}\left\{ 1-\E\left(\frac{M_0}{H_1+M_0}\right)^2 + \E\left(\frac{M_0}{H_0+1+M_0}\right)^2 \left(\frac{N_0}{N_1M_0}-1\right) \right\}
        \right].
\end{align*}}
It is obvious that $\E\left(\frac{M_0}{H_1+M_0}\right)^2\leq1$ , and we have $N_0>N_1M_0$ by assumption.
Hence,
\[
N\{\var(\hat{\tau}_{0}^{t}) - \var(\hat{\tau}^{t})\} >0.
\]
Then the result follows.\qed

\subsection{Proof of Theorem \ref{thm:2}}
We consider each of the terms in (\ref{eqt:decomposition}) separately.
For the first term, by the law of large numbers $\bar{\tau} (X)\rightarrow \tau$ in probability.
For the weighted average of the residual $R $, we have $N \E\{(R )^2\}=\E[\{1+K(i)/M_0\}^2\sigma^2(d,x)]=O(1)$, which is finite, hence, $R  = O_p(N^{-1/2}
) = o_p(1)$.
And for the bias term, by Theorem \ref{thm:1}, $B  = O_p(N^{-1/k}) = o_p(1)$.
We thus prove the first part of the theorem.
The proof of the second part follows the same argument and is therefore omitted.\qed

\subsection{Proof of Theorem \ref{thm:3}}
Since the second part of the theorem follows the same argument as the first part, we only prove part (i).
First, $(V^{\tau(X)})^{-1/2}N^{1/2}\{\bar{\tau}(X)-\tau\} \rightarrow  \Normal(0, 1)$ in distribution by central limit theorem.
Second, since $\E[\{1+K(i)/M_0\}^4]$ is uniformly bounded, we then obtain
\[
\frac{N^{1/2}\sum_{i=1}^NR_{i}}{[\sum_{i=1}^N\{1+K(i)/M_0\}^2\sigma^2(D_i,X_i)]^{1/2}}=(V^R)^{-1/2}N^{1/2}R \rightarrow \Normal(0, 1)
\] in distribution in a similar manner to Proof of Theorem~4 of \citet{abadie2006large} using a Lindeberg-Feller central limit theorem.
In addition, $(V^R)^{-1/2}N^{1/2}R$ and $(V^{\tau(X)})^{-1/2}N^{1/2}\{\bar{\tau}(X)-\tau\}$ are asymptotically independent.
Therefore, the convergence to $\Normal(0, 1)$ of both $(V^R)^{-1/2}N^{1/2}R$ and $(V^{\tau(X)})^{-1/2}N^{1/2}\{\bar{\tau}(X)-\tau\}$, boundedness of $V^R$ and $V^{\tau(X)}$, and boundedness of $V^R$ away from zero imply that
\[\{V^R + V^{\tau(X)}\}^{-1/2}
N^{1/2}
(\hat{\tau}  - B  - \tau) \rightarrow  \Normal(0, 1)\] in distribution.
Then the desired results follow. \qed

\subsection{Proof of Theorem \ref{thm:5}}
We only prove the first part since part (ii) of the theorem follows the same argument as part (i).
Since it is proven in Section \ref{Sec:Thm1Proof} that $\sup_{j \in \mathcal{J}(i)}\|X_{i}-X_{j}\|$ is bounded,
it follows from Lemma~A.2 of \citet{abadie2011bias} that
\[
\max_{i=1,\ldots,N} |\hat{\mu}(D_i,X_i)-\hat{\mu}(D_i,X_{j})-\{\mu(D_i,X_i)-\mu(D_i,X_{j})\}|
    =  o_p(N^{-1/2}),
\]
where $j\in \mathcal{J}(i)$.
The difference $|\hat{B} - B|$
can be written as
{\small
\begin{align*}
|\hat{B} -& B|\\
\leq& \frac{1}{N}\sum^N_{i=1}\frac{1}{M(i)}\sum_{j\in \mathcal{J}(i)}| D_i\{\hat{\mu}(0,X_i)-\hat{\mu}(0,X_{j})\}-(1-D_i)\{\hat{\mu}(1,X_i)-\hat{\mu}(1,X_{j})\}\\
& -[D_i\{\mu(0,X_i)-\mu(0,X_{j})\}-(1-D_i)\{\mu(1,X_i)-\mu(1,X_{j})\}]|
\\
    \leq &  \max_{\substack{i=1,\ldots,N\\j\in \mathcal{J}(i)}}\sum_{d=0,1} |\hat{\mu}(d,X_i)-\hat{\mu}(d,X_{j})-\{\mu(d,X_i)-\mu(d,X_{j})\}|\\
    = &  o_p(N^{-1/2}).
\end{align*}}
Then the desired results follow. \qed

\subsection{Proof of Theorem \ref{thm:6}}
First notice that
{\small\begin{align*}
& \left|\frac{1}{N} \sum_{i=1}^N\left\{1+\frac{K(i)}{M_0}\right\}^2 \{\sigma^2(D_i,X_i)-\hat{\sigma}^2(D_i,X_i) \}\right| \\
& \qquad\leq\left|\frac{1}{N} \sum_{i=1}^N\left(1+\frac{K(i)}{M_0}\right)^2 [\sigma^2(D_i,X_i)-\E\{\hat{\sigma}^2(D_i,X_i) \mid X_{1:N}, D_{1:N}\}]\right| \\
& \qquad\qquad+\left| \frac{1}{N} \sum_{i=1}^N\left\{1+\frac{K(i)}{M_0}\right\}^2  [\E\{\hat{\sigma}^2(D_i,X_i) \mid X_{1:N}, D_{1:N}\}-\hat{\sigma}^2(D_i,X_i)]\right|.
\end{align*}}
Given Lemma \ref{lem:1.1} that all moments of $K(i)$ are bounded uniformly in $N$,
it follows from Proof of Theorem~6 of \citet{abadie2002simple} that both terms on the right-hand side of the above inequality converge to zero.
And thus \begin{equation}\label{eqt:VR}
\frac{1}{N} \sum_{i=1}^N\left\{1+\frac{K(i)}{M_0}\right\}^2 \hat{\sigma}^2(D_i, X_i) \rightarrow V^R
\end{equation}
in probability.
Now, consider
\begin{align*}
\frac{1}{N} \sum_{i=1}^N\{\Tilde{Y}_i(1)-\Tilde{Y}_i(0)-\hat{\tau}\}^2 & =\frac{1}{N} \sum_{i=1}^N\{\Tilde{Y}_i(1)-\Tilde{Y}_i(0)-\tau\}^2-(\hat{\tau}-\tau)^2 \\
& =\frac{1}{N} \sum_{i=1}^N\{\Tilde{Y}_i(1)-\Tilde{Y}_i(0)-\tau\}^2+o_p(1) .
\end{align*}
We decompose
{\scriptsize
\begin{align*}
& \frac{1}{N} \sum_{i=1}^N\left\{\Tilde{Y}_i(1)-\Tilde{Y}_i(0)-\tau\right\}^2 \\
& =\frac{1}{N} \sum_{i=1}^N\left[\left(2 D_i-1\right)\left\{\frac{1}{M(i)}\sum_{j\in \mathcal{J}(i)} \mu(D_i,X_i)-\mu(1-D_i,X_{j})\right\}-\tau\right]^2+\frac{1}{N} \sum_{i=1}^N\left\{\epsilon_i-\frac{1}{M(i)}\sum_{j\in \mathcal{J}(i)} \epsilon_{j}\right\}^2 \\
& +\frac{2}{N} \sum_{i=1}^N\left[\left(2 D_i-1\right)\left\{\frac{1}{M(i)}\sum_{j\in \mathcal{J}(i)} \mu(D_i,X_i)-\mu(1-D_i,X_{j})\right\}-\tau\right]\left(2 D_i-1\right)\left\{\epsilon_i-\frac{1}{M(i)}\sum_{j\in \mathcal{J}(i)} \epsilon_{j}\right\}.
\end{align*}
}
Since the sample maximum of the norms of the matching discrepancies $\left\|X_i-X_{j}\right\|$ is $o_p(1)$, and the regression functions $\mu(d,x)$ are Lipschitz, for $d = 0,1$, we follow a similar argument to the proof of Theorem~7 of \citet{abadie2006large} to obtain
\begin{align*}
& \frac{1}{N} \sum_{i=1}^N\left[(2 D_i-1)\left\{\frac{1}{M(i)}\sum_{j\in \mathcal{J}(i)} \mu(D_i,X_i)-\mu(1-D_i,X_{j})\right\}-\tau\right]^2 \\
& =\frac{1}{N} \sum_{i=1}^N\{\mu(1,X_i)-\mu(0,X_i)-\tau\}^2+o_p(1)
\end{align*}
for the first term on the right-hand side of the above equation.
Also similar to Proof of Theorem~7 of \citet{abadie2006large}, we have
\[
\frac{1}{N} \sum_{i=1}^N\left\{\epsilon_i-\frac{1}{M(i)} \sum_{j\in \mathcal{J}(i)} \epsilon_{j}\right\}^2-\frac{1}{N} \sum_{i=1}^N\left\{1+\frac{K(i)}{M_0^2}\right\} \sigma^2(D_i, X_i)=o_p(1)
\]
and
\begin{align*}
\frac{1}{N} \sum_{i=1}^N\{\Tilde{Y}_i(1)-\Tilde{Y}_i(0)-\hat{\tau}\}^2 & =\frac{1}{N} \sum_{i=1}^N\left\{\mu_1\left(X_i\right)-\mu_0\left(X_i\right)-\tau\right\}^2 \\
& \qquad +\frac{1}{N} \sum_{i=1}^N\left\{1+\frac{K(i)}{M_0^2}\right\} \sigma^2(D_i, X_i)+o_p(1)
\end{align*}
for the second and the last terms on the right-hand side of the equation, respectively.
Similar to the convergence (\ref{eqt:VR}), it can be obtained that
\[
\left|\frac{1}{N} \sum_{i=1}^N\left\{1+\frac{K(i)}{M_0^2}\right\} \sigma^2(D_i, X_i)-\frac{1}{N} \sum_{i=1}^N\left\{1+\frac{K(i)}{M_0^2}\right\} \hat{\sigma}^2(D_i, X_i)\right|=o_p(1).
\]
Therefore,
\begin{equation}\label{eqt:Vtau}
\frac{1}{N} \sum_{i=1}^N\left\{\Tilde{Y}_i(1)-\Tilde{Y}_i(0)-\hat{\tau}\right\}^2-\frac{1}{N} \sum_{i=1}^N\left\{1+\frac{K(i)}{M_0^2}\right\} \hat{\sigma}^2(D_i, X_i) \rightarrow V^{\tau(X)}
\end{equation}
in probability.
Since
\begin{align*}
\hat{V} &  =\frac{1}{N} \sum_{i=1}^N\{\Tilde{Y}_i(1)-\Tilde{Y}_i(0)-\hat{\tau}\}^2 \\
&\qquad +\frac{1}{N} \sum_{i=1}^N\left[\left\{\frac{K(i)}{M_0}\right\}^2+\left(\frac{2 M-1}{M_0}\right)\left\{\frac{K(i)}{M_0}\right\}\right] \hat{\sigma}^2(D_i, X_i)\\
&= \frac{1}{N} \sum_{i=1}^N\{\Tilde{Y}_i(1)-\Tilde{Y}_i(0)-\hat{\tau}\}^2 + \frac{1}{N} \sum_{i=1}^N\left\{1+\frac{K(i)}{M_0}\right\}^2 \hat{\sigma}^2(D_i, X_i) \\
& \qquad -\frac{1}{N} \sum_{i=1}^N\left\{1+\frac{K(i)}{M_0^2}\right\} \hat{\sigma}^2(D_i, X_i),
\end{align*}
the convergences (\ref{eqt:VR}) and (\ref{eqt:Vtau}) imply that
\begin{align*}
\hat{V} \rightarrow V^R+V^{\tau(X)}
\end{align*}
in probability,
which finishes the proof for the first part of the theorem.
The proof for the second part is similar to that of part one and is omitted. \qed

\section{Simulation results}

In our simulation designs,
the average treatment effects in this simulation design are set as $\tau = \tau^t = 1$.
The parameter $c$ controls the ratio of the treated.
Since matching estimators have been used when the interest is in $\tau^t$, and there are more controls than the treated, we adjust $c$ to fix the ratio of the treated roughly at $0.3$ for all cases.
For example, letting $c=0.4$ will lead to a treated ratio of $0.3$ for all single covariate cases.
For multivariate cases, we fix $c$ at $0.45$ for Cases 2(a), 2(b), 2(d), and 2(e) and at $0.55$ for Case 2(c);
see Remark \ref{rem:otherc} for more discussions.
For each case, we set the sample size as
$N = 100$, the number of fixed matches as $M_0 = 1, 2, 3, 4, 5$, and the number of Monte Carlo replications as $2^{12}$.

\begin{figure}[t!]
	\begin{center}
		\includegraphics[width=.99\textwidth]{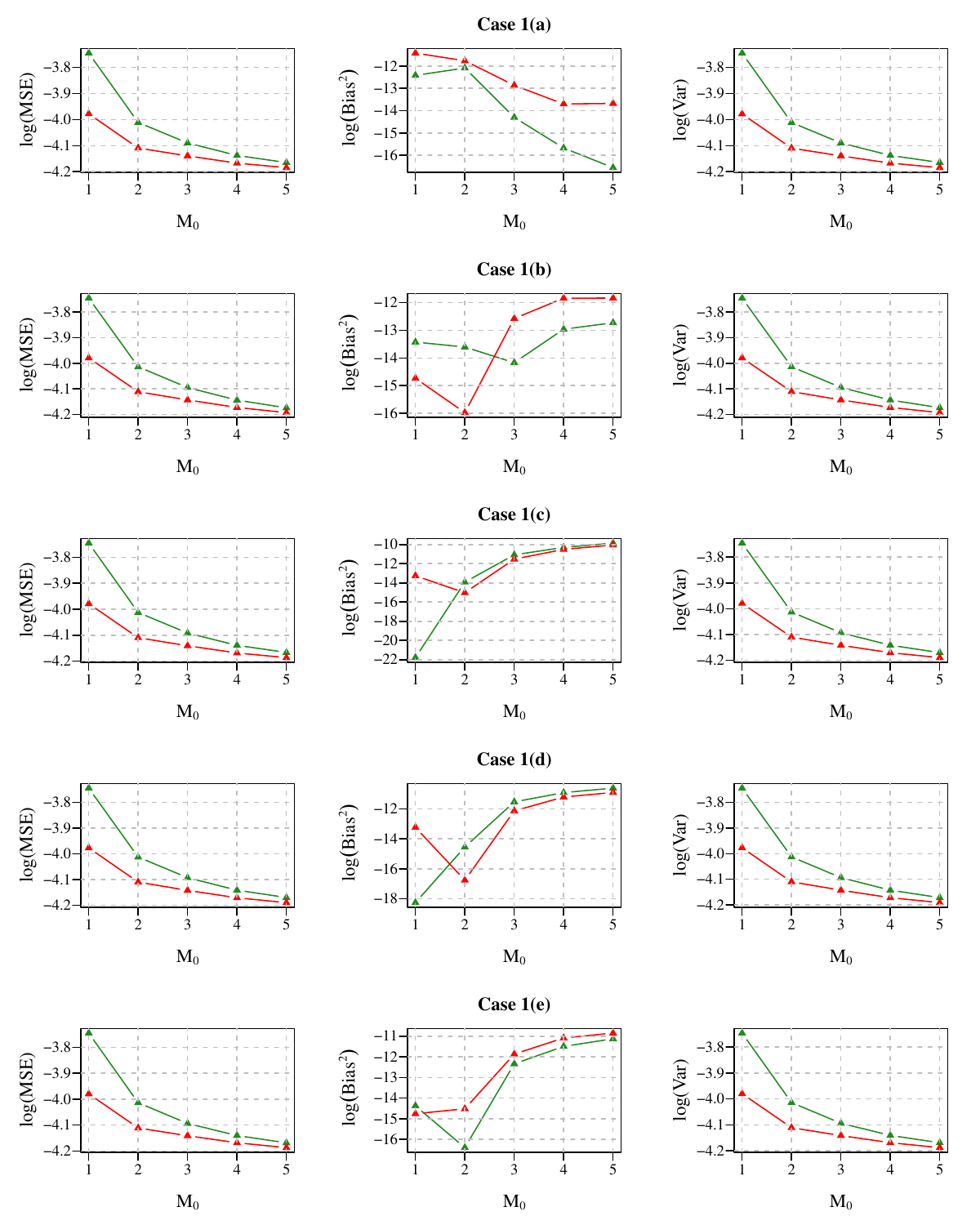}
		\caption{A graph comparing the performance between $\hat{\tau}_{0}^{t}$ (green triangles) and $\hat{\tau}^{t}$ (red triangles) when $X$ is a scalar.}
		\label{fig:scalar_X}
	\end{center}
\end{figure}

Five cases with a single covariate are considered.
In particular, we let $X \sim\Beta(1.2, 1.2)$,
where $\Beta(\alpha, \beta)$ denotes the beta distribution with parameters $\alpha, \beta>0$.
For the mean function $m(x)$, we consider five non-linear curves presented in Table \ref{tb:m}.
We set $P(x) = 0.15 + 0.7 x$ which assigns a unit to the treated group with higher probability
if its corresponding covariate is of a greater value.
The simulation results are in Figure \ref{fig:scalar_X}.

\begin{figure}[t!]
	\begin{center}
		\includegraphics[width=.99\textwidth]{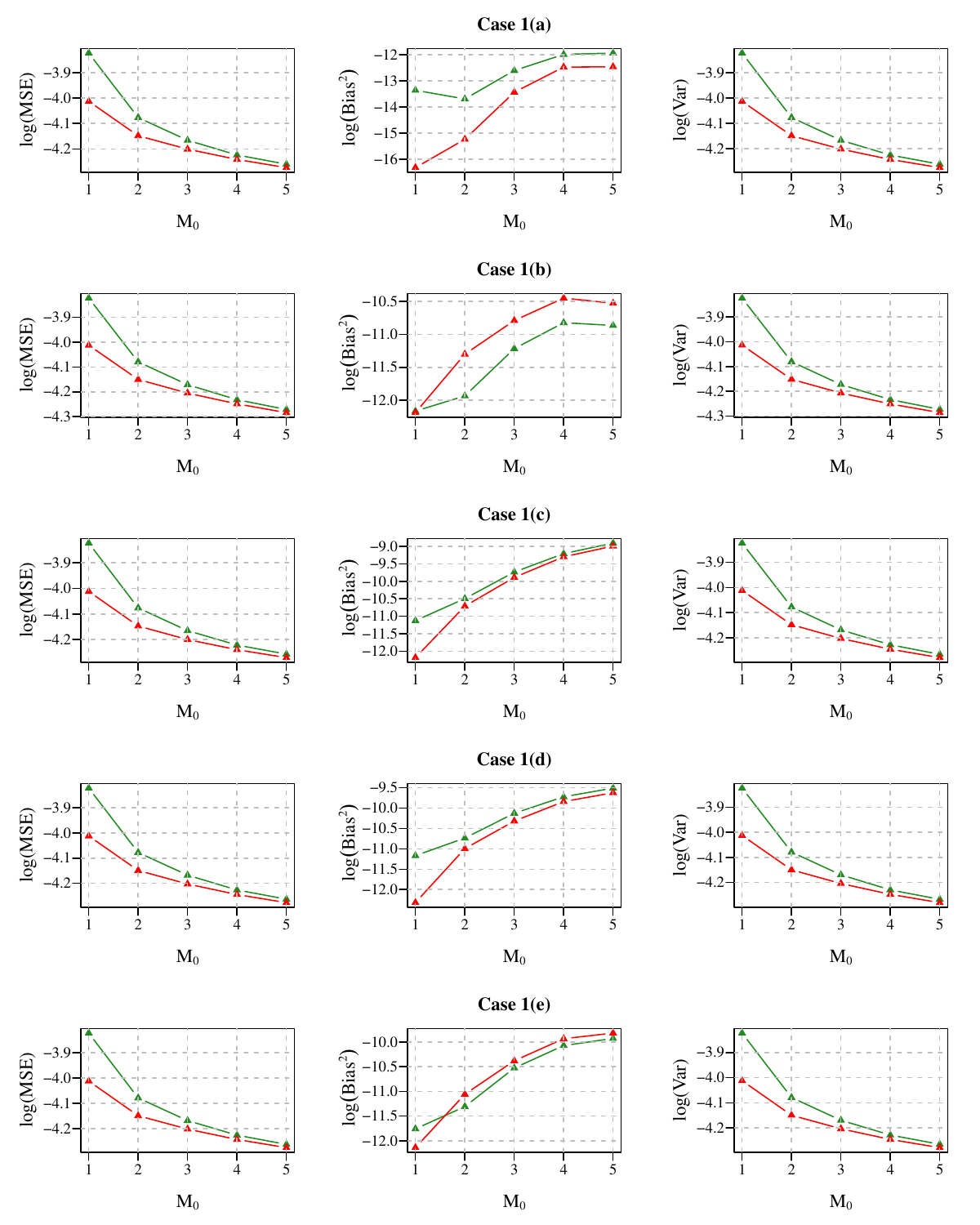}
		\caption{A graph comparing the performance between $\hat{\tau}_{0}^{t}$ (green triangles) and $\hat{\tau}^{t}$ (red triangles) when $X$ is a scalar and $c=0.3$.}
		\label{fig:c0.3}
	\end{center}
\end{figure}

\begin{remark}\label{rem:otherc}
In a preliminary simulation study, we consider different treated ratio and find that results are similar overall.
For example, if we choose a smaller $c$, the treated ratio will be larger.
But as long as we have more treated units than control units, our conclusion will not change with the choice of $c$.
Figure \ref{fig:c0.3} shows the results with the same simulation setting except that $c=0.3$ for all cases.
The same conclusions are obtained.
\end{remark}

\begin{table}[t!]
\caption{Simulation designs for $m(x)$.}
\centering
\renewcommand{\arraystretch}{1.1}
\begin{tabular}{|ll|}\hline
Cases & $m(x)$ \\ \hline
1(a) & $0.1 + z/2 + \exp(-200(z - 0.7)^2 )/2$\\
1(b) & $0.8 - 2(z - 0.9)^2 - 5(z - 0.7)^3 - 10(z - 0.6)^{10}$\\
1(c) &  $0.2 + \sqrt{1 - z} - 0.6(0.9 - z)^2$\\
1(d) &  $0.2 + \sqrt{1 - z} - 0.6(0.9 - z)^2 - 0.1z \cos(30z)$\\
1(e) &  $0.4 + 0.25 \sin(8z - 5) + 0.4 \exp(-16(4z-2.5)^2)$\\ \hline
\end{tabular}
\label{tb:m}
\end{table}

We also consider cases with multiple covariates $X_i^\T=(X_{i1},\ldots,X_{i6})$.
In particular, we set the number of covariates at $6$ with one covariate being discretely distributed, i.e., $k=5$.
For the first three covariates, $(X_{i1},X_{i2},X_{i3})^\T$
is generated from a multivariate normal distribution
with mean $\mu_x\in\mathbb{R}^3$,
$\var(X_{i1})=2$, $\var(X_{i2})=\var(X_{i3})=\cov(X_{i1},X_{i2})=1$,
$\cov(X_{i1},X_{i3})=-1$, and $\cov(X_{i2},X_{i3})=-0.5$.
For the remaining three covariates,
we generate $X_{i4} \sim \Unif(-3, 3)$, $X_{i5} \sim \chi^2_1$, and
$X_{i6} \sim \Bern(0.5)$, where
$\chi^2_1$ denotes the chi-squared distribution with one degree of freedom.
Suppose that the mean function admits the form $m(x)=\gamma^\T g(x)$,
where $x=(x_1,x_2, \ldots, x_6)^{\T}$,
$\gamma\in\mathbb{R}^h$ is a $h$-vector of coefficients,
$g:\mathbb{R}^k\rightarrow \mathbb{R}^h$ is some link function, and
$h\in\mathbb{N}$.
The propensity score $P(x) = x^\T\eta$,
where $\eta\in\mathbb{R}^k$ is a vector of coefficient.
Five cases for $\mu_x$, $\eta$, $g$ and $\gamma$ are considered; see Table \ref{tb:3cases}.

\begin{table}[t!]
\caption{The values of parameters used in the simulation experiments.}
\centering
\renewcommand{\arraystretch}{1.1}
\begin{tabular}{|lll|}\hline
Cases & Parameters & Values \\ \hline
2(a) & $\mu_x$ & $(0, 0, 0)^{\T}$ \\
    &              $\eta$ & $0.7\times(0, 1, 1, 1, -2, 1, 0.5)^{\T}$\\
    &              $g(x)$ & $(1, x_{1:2}^{\T}, (x_3/2), \exp(-200(x_4-0.7)^2), x_{5:6}^{\T})^{\T}$\\
    &              $\gamma$ & $(-3, 2, 1, 1, 2, 2, 2)^{\T}$ \\ \hline
2(b) & $\mu_x$ & $(0, 0, 0)^{\T}$\\
   &               $\eta$ & $0.7\times(0, 1, 1, 1, -2, 1, 0.5)^{\T}$, \\
   &               $g(x)$ & $(1, x_{1}, x_2^4, x_3, \sqrt{3-x_4}, x_{5:6}^\T)^{\T}$,\\
   &               $\gamma$ & $(-3, 2, 1, 1, 2, -1, 1)^{\T}$ \\ \hline
2(c) & $\mu_x$ & $(-0.5, 0, 0)^{\T}$ \\
   &         $\eta$ & $0.7\times(0, -1, 1.5, 1, -0.5, 1, 0.5)^{\T}$\\
   &         $g(x)$ & $(1, x_{1:6}^{\T})^{\T}$\\
    &        $\gamma$ & $(-3, 2, 1, 1, 2, 2, 2)^{\T}$\\ \hline
2(d) & $\mu_x$ & $(0, 0, 0)^{\T}$\\
   &               $\eta$ & $0.7\times(0, 1, 1, 1, -2, 1, 0.5)^{\T}$, \\
   &               $g(x)$ & $(1, x_1, (0.2-x_2)^3, x_{3:6}^{\T}$,\\
   &               $\gamma$ & $(-3, 2, 1, 1, 2, 2, 2)^{\T}$ \\ \hline
2(e) & $\mu_x$ & $(0, 0, 0)^{\T}$\\
   &               $\eta$ & $0.7\times(0, 1, 1, 1, -2, 1, 0.5)^{\T}$, \\
   &               $g(x)$ & $(1, x_{1}, x_2^4, (1.5-x_3)^3, \sqrt{3-x_4}, (x_{5}-0.9)^2, x_6)^{\T}$,\\
   &               $\gamma$ & $(-3, 2, 1, 1, 2, 2, 2)^{\T}$ \\ \hline
\end{tabular}
\label{tb:3cases}
\end{table}

Since the dimension of covariate is $k>1$,
the bias term is non-negligible and a bias adjustment is needed.
The bias can be corrected by the ordinary least-squares estimator via a linear regression model, which is a special case of the power series estimator.
As it is implemented by \cite{abadie2004implementing}, we use the weighted linear regression for correcting the bias of our proposed estimators, i.e., $\hat{\mu}(d,x) = \hat{\alpha}_d + x^\T\hat{\beta}_d$ for $d=0,1$, where $(\hat{\alpha}_d,\hat{\beta}_d)=\argmin_{(\alpha_d,\beta_d)}\sum_{i:D_i=d}K(i)/M_0(Y_i-\alpha_d - X_i^\T\beta_d)^2$.
The weight $K(i)/M_0=\sum^N_{\ell=1}\mathbb{1}\{i \in \mathcal{J}(\ell)\} /M(\ell)$ is the number of times unit $i$ is used as a match for all units $\ell$, each time weighted by the total number of matches for unit $\ell$.
Notice that here we do the matching by Mahalanobis distance which is the vector norm $\lVert x\rVert_A = (x^\prime Ax)^{1/2}$, where the $k \times k$ matrix $A$ is chosen to be the inverse of the sample covariance matrix of the covariates, corresponding to the Mahalanobis metric:
\[
A = \left\{\frac{1}{N}\sum_{i=1}^N(X_i-\bar{X})(X_i-\bar{X})^\T\right\}^{-1},
\]
where $\bar{X}=\sum_{i=1}^NX_i/N$.
We also include in the comparison the genetic matching estimator \citep{diamond2013genetic}, which uses a state-of-art iterative algorithm to maximize a criterion related to covariate balance and perform the nearest neighbour matching using the scaled generalized Mahalanobis distance.
However, the genetic matching has a few tuning parameters that slow the estimation for better results.
For instance, a higher value of the tuning parameter called `pop.size' of the genetic matching slower the estimation for better results.
For fairness, we set `pop.size=20' to make the computational time roughly the same for all estimators.
If we allow the computational time to be longer for the genetic matching estimator, it is likely that it will perform better in terms of the covariate balance,
but at the cost of efficiency.

\begin{figure}[t!]
	\begin{center}
		\includegraphics[width=.99\textwidth]{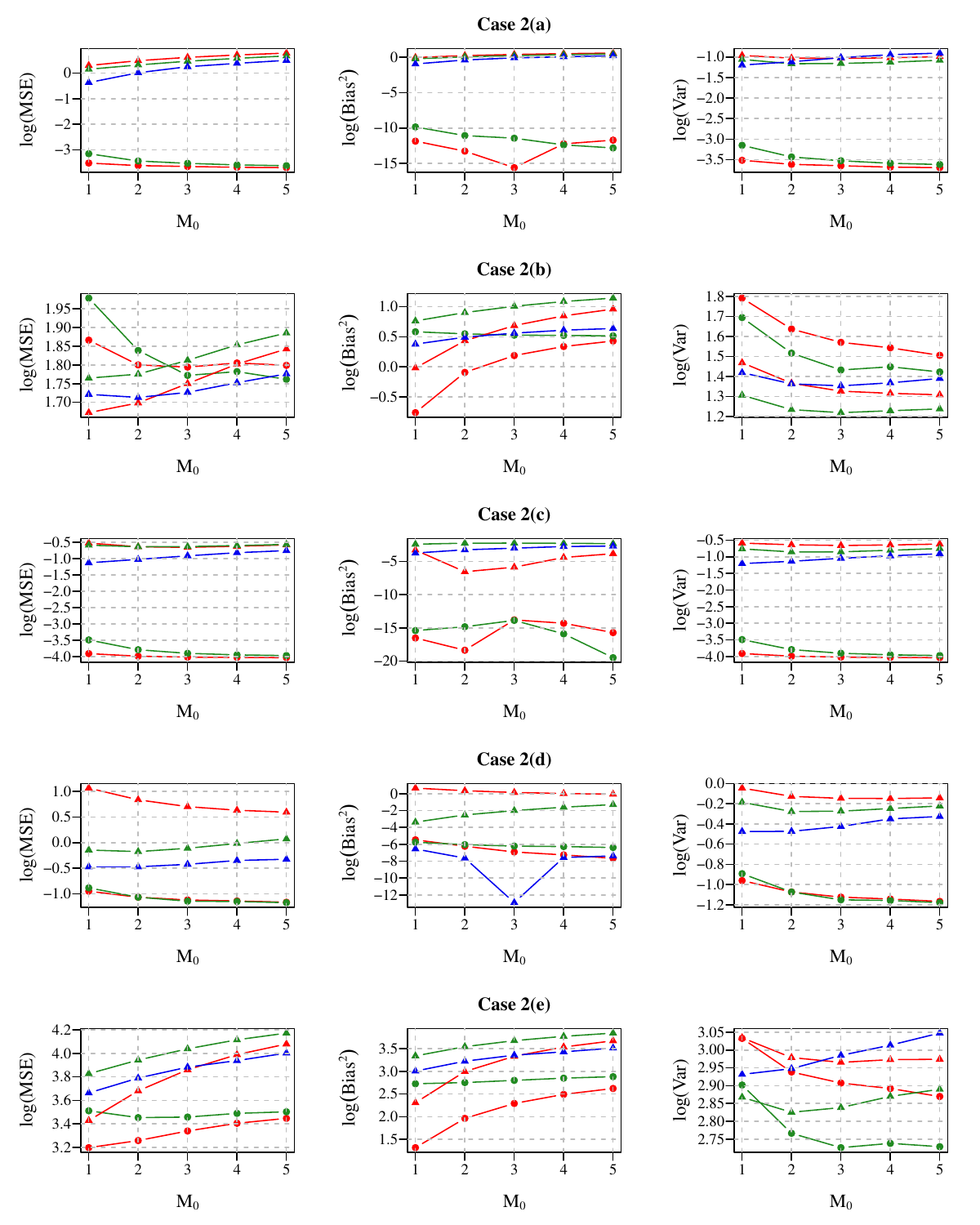}
		\caption{A graph comparing the performance among $\hat{\tau}_{0}^{t}$ (green triangles), $\hat{\tau}^{t}$ (red triangles), $\hat{\tau}^{t}_{0,\bc}$ (green circles), $\hat{\tau}^{t}_{\bc}$ (red circles), and the genetic matching estimator $\hat{\tau}^{t}_{gen}$ (blue triangles) when $X$ is multi-dimensional.}
		\label{fig:simulation_all}
	\end{center}
\end{figure}

\begin{figure}[t!]
	\begin{center}
		\includegraphics[width=.99\textwidth]{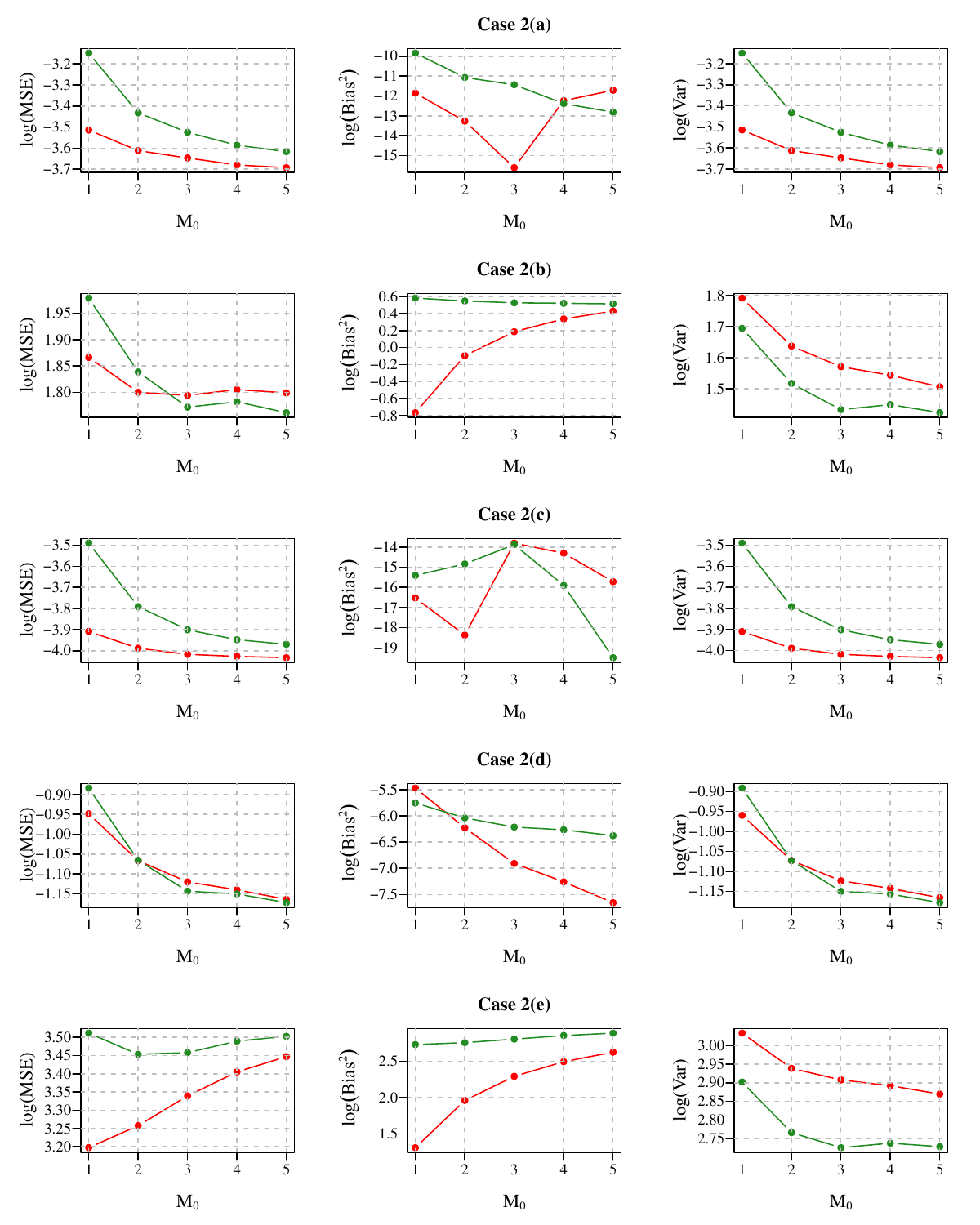}
		\caption{A graph comparing the performance between $\hat{\tau}^{t}_{0,\bc}$ (green circles) and $\hat{\tau}^{t}_{\bc}$ (red circles) when $X$ is multi-dimensional.}
		\label{fig:multi_dimentional_X}
	\end{center}
\end{figure}

Figure \ref{fig:simulation_all} is the performance comparison when $X$ is multi-dimensional.
It is found that the mean squared errors of $\hat{\tau}^t_{\bc}$ and $\hat{\tau}^{t}_{0,\bc}$ are much lower compared with others.
Hence, we extract the results for $\hat{\tau}^t_{\bc}$ and $\hat{\tau}^{t}_{0,\bc}$ in Figure \ref{fig:multi_dimentional_X} for better comparison.
Interpreting the results from multi-dimensional cases is difficult due to the bias estimation. Using more data points allows a better fit but meanwhile introduces more bias if they are not good matches.

Intuitively, a complicated mean function requires more data for a better bias estimation, which is the case for Case 2(d) in Figure \ref{fig:multi_dimentional_X}.
Therefore, we may discover a decrease in bias with increasing $M_0$ without rematching. However, the rematching estimator performs better in terms of bias across $M_0$ since rematching can provide extra information for the underlying mean function. In fact, the rematching estimator uses more information with more weights given to data matched with high frequency.

Case 2(e) in Figure \ref{fig:multi_dimentional_X} shows a decreasing trend in bias possibly because the mean function is the most complex and increasing $M_0$ introduces more bias which is hard to correct since the outcome model is not well-estimated even more data are used.
Our rematching estimator performs better because it avoids repeatedly reusing matches that are too far away and hence does not introduce much bias when using more information.

\begin{table}[t!]
\caption{Coverage rates of the 95\% and the 90\% confidence intervals.}
\centering
\renewcommand{\arraystretch}{1.1}
\setlength{\tabcolsep}{3.0pt}
\begin{tabular}{|cccccccccccc|}\hline
   & &  \multicolumn{2}{c}{$M_0=1$}  &\multicolumn{2}{c}{$M_0=2$} & \multicolumn{2}{c}{$M_0=3$} & \multicolumn{2}{c}{$M_0=4$} & \multicolumn{2}{c|}{$M_0=5$} \\
CI & Cases & 95\% & 90\% & 95\% & 90\% & 95\% & 90\% & 95\% & 90\% & 95\% & 90\%\\\hline
$\hat{C}^t$ &1(a) &0.939 &0.885 &0.938 &0.889 &0.935 &0.890 &0.936 &0.884 &0.938 &0.885\\
&1(b) &0.939 &0.885 &0.938 &0.892 &0.935 &0.891 &0.935 &0.885 &0.937 &0.882\\
&1(c) &0.939 &0.885 &0.939 &0.888 &0.936 &0.891 &0.937 &0.885 &0.940 &0.882\\
&1(d) &0.937 &0.885 &0.939 &0.890 &0.935 &0.891 &0.939 &0.886 &0.939 &0.885\\
&1(e) &0.940 &0.883 &0.938 &0.889 &0.937 &0.889 &0.936 &0.887 &0.937 &0.883\\\hline
$\hat{C}_0^t$ &1(a) &0.940 &0.888 &0.942 &0.891 &0.938 &0.884 &0.939 &0.886 &0.937 &0.883\\
&1(b) &0.941 &0.886 &0.941 &0.891 &0.938 &0.885 &0.937 &0.884 &0.937 &0.886\\
&1(c) &0.941 &0.888 &0.941 &0.890 &0.938 &0.888 &0.939 &0.884 &0.939 &0.882\\
&1(d) &0.939 &0.887 &0.943 &0.891 &0.938 &0.885 &0.938 &0.886 &0.939 &0.885\\
&1(e) &0.941 &0.887 &0.941 &0.891 &0.939 &0.884 &0.938 &0.883 &0.937 &0.883\\\hline
\end{tabular}
\label{tb:covRate}
\end{table}

We also evaluate the performance of the asymptotic variance estimator proposed.
The implementation of the variance estimation follows a similar manner to the software documented in \cite{abadie2004implementing}.
We consider the simulation Cases 1(a)--(e).
The $100(1-\alpha)\%$ confidence intervals for the rematching estimator $\hat{\tau}^t$ and the simple matching estimator $\hat{\tau}_0^t$ are computed as
\begin{align*}
\hat{C}^t &= \left[\hat{\tau}^t-q_{1-\alpha/2}(\hat{V}^t)^{1/2},\quad \hat{\tau}^t+q_{\alpha/2}(\hat{V}^t)^{1/2} \right], \\
\hat{C}_0^t &= \left[\hat{\tau}_0^t-q_{1-\alpha/2}(\hat{V}_0^t)^{1/2},\quad \hat{\tau}_0^t+q_{\alpha/2}(\hat{V}_0^t)^{1/2} \right],
\end{align*}
respectively, where $q_{1-\alpha/2}$ and $q_{\alpha/2}$ are the $(1-\alpha/2)$th and $(\alpha/2)$th quantiles of the standard normal distribution, and $\hat{V}_0^t$ is defined in  Section \ref{sec:var_est}.
For each estimator, we report the coverage rates of the 95\% and the 90\% confidence intervals in Table \ref{tb:covRate}.
It is found that the coverages of confidence intervals are very close to the nominal levels across different values of the number of fixed matches and different simulation cases,
which means the matching-based variance estimator works well for both estimators.

\section{National Supported Work data}

In particular, we use nine variables: RE78 (earnings in 1978), treatment indicator (1 if treated, 0 if not treated), age, education, Black (1 if black, 0 otherwise), Hispanic (1 if Hispanic, 0 otherwise), married (1 if married, 0 otherwise), RE74 (earnings in 1974), RE75 (earnings in 1975), where RE78 is the outcome variable and the last seven variables are covariates.
The dataset we use contains three groups including the experimental treated and control groups from a randomized evaluation of the NSW program, and a subset of the nonexperimental control group from the Panel Study of Income Dynamics (PSID).
According to \cite{dehejia1999causal}, there is a lack of overlap in the pretreatment covariates of the PSID data and the experimental treated group.
A subset of the nonexperimental control group named PSID-2 is therefore extracted by Lalonde from PSID to bridge the gap between the treated and the nonexperimental control groups.

We compute experimental matching estimates using the experimental treated and control groups and nonexperimental matching estimates using the experimental treated group and PSID-2.
For each pair of the treated and control groups, we implement matching estimators based on the Mahalanobis distance and estimate their standard errors.
The estimates include matching estimates without bias correction and bias-adjusted matching estimates, first by the simple matching and then by the matching-and-rematching procedure.
For the bias correction, we estimate the regression function $\hat{\mu}(0, x)$ by the ordinary least squares of the linear regression on all covariates.
The computation details for finding matches, doing regression adjustments and estimating standard errors, follow a similar manner to the software documented in \cite{abadie2004implementing}.
For the number of fixed matches $M_0$, we consider the cases $M_0 = 1, 4, 16, 64, N_0$, where $N_0$ is 260 and 253, respectively, for the experimental and the nonexperimental controls.
The matching estimates reduce to difference-in-means estimates when $M_0=N_0$.

\section{Results from Abadie and Imbens's works}
\subsection{Consistency and asymptotic normality}
\cite{abadie2006large} showed that $\hat{\tau}_{0}$ is consistent and asymptotically normal under Assumptions \ref{a:1}, \ref{a23}, and \ref{a:4}, i.e.,
\begin{equation}
    (V_{0})^{-1/2}N^{1/2} (\hat{\tau}_{0} - B_{0} - \tau )
\rightarrow  \Normal(0, 1) \nonumber
\end{equation} in distribution, where $B_{0}$ and $V_{0}$ are asymptotic
bias and variance terms, respectively, defined as
\begin{align*}
B_{0} &= \frac{1}{N}\sum_{i=1}^N
(2D_i - 1)
\left[\frac{1}{M_0}\sum_{j\in \mathcal{J}_{M_0} (i)}
\{\mu(1 - D_i, X_i)
- \mu(1 - D_i, X_j)\}\right],\\
V_{0} &= V^{R}_0+V^{\tau(X)},\\
V^{R}_0 &= \frac{1}{N}\sum_{i=1}^N \left\{1 + \frac{K_{M_0}(i)}{M_0}\right\}^2
\sigma^2(D_i, X_i),\\
V^{\tau(X)} &= \E\{\mu(1, X) - \mu(0, X) - \tau\}^2.
\end{align*}
Under Assumptions \ref{a:1}, \ref{a2p3p}, and \ref{a:4}, similar results hold for its counterpart for the treated population as follows:
\begin{equation}
    (V^t_{0})^{-1/2}N_1^{1/2}(\hat{\tau}^t_{0} - B^t_{0} - \tau^t )
\rightarrow  \Normal(0, 1) \nonumber
\end{equation}in distribution, where $B^t_{0}$ and $V^t_{0}$ are asymptotic
bias and variance terms, respectively, defined as
\begin{align*}
B^t_{0} &= \frac{1}{N_1}\sum_{i=1}^N
D_i
\left[\frac{1}{M_0}\sum_{j\in \mathcal{J}_{M_0} (i)}
\{\mu(0, X_i)
- \mu(0, X_j)\}\right],\\
V^t_{0} &= V^{R,t}_0+V^{\tau(X),t},\\
V^{R,t}_0 &= \frac{1}{N_1}\sum_{i=1}^N \left\{D_i-(1-D_i) \frac{K_{M_0}(i)}{M_0}\right\}^2
\sigma^2(D_i, X_i),\\
V^{\tau(X),t} &= \E\left[\{\mu(1, X) - \mu(0, X) - \tau^t\}^2\mid D=1\right].
\end{align*}

\subsection{Bias correction}
\cite{abadie2011bias} proposed to estimate $B_{0}$ and $B_{0}^t$ by
\begin{align*}
    \hat{B}_0 &= \frac{1}{N} \sum_{i=1}^N
    \frac{(2D_i - 1)}{M_0}\sum_{j\in \mathcal{J}_{M_0} (i)}\{ \hat{\mu}(1 - D_i, X_i) - \hat{\mu}(1 - D_i, X_j)\},\\
    \hat{B}_0^t &= \frac{1}{N_1} \sum_{i=1}^N
    \frac{D_i}{M_0}\sum_{j\in \mathcal{J}_{M_0} (i)}\{ \hat{\mu}(0, X_i) - \hat{\mu}(0, X_j)\},
\end{align*}
respectively, where $\hat{\mu}(d, x)$ is a non-parametric series regression estimator that satisfies Assumption \ref{a2}.
By Theorem~2 of \citet{abadie2011bias},
\[
N^{1/2}(B_{0} - \hat{B}_0)\rightarrow 0 \]
in probability, and
\[\{V^{R}_0+V^{\tau(X)}\}^{-1/2}N^{1/2}(\hat{\tau}_{0} - \hat{B}_0 - \tau )
\rightarrow \Normal(0, 1)
\]
in distribution,
and by Theorem~2$^\prime$ of \citet{abadie2011bias},
\[
N_1^{1/2}(B^t_{0} - \hat{B}^t_0)\rightarrow 0 \]
in probability, and
\[\{V^{R,t}_0+V^{\tau(X),t}\}^{-1/2}N_1^{1/2}(\hat{\tau}^t_{0} - \hat{B}^t_0 - \tau^t )
\rightarrow \Normal(0, 1)
\]
in distribution,
meaning that the bias adjustment does not affect the asymptotic variances.

\subsection{Variance estimation}\label{sec:var_est}
As suggested by \cite{abadie2006large}, the asymptotic variances can be consistently estimated by
\begin{align*}
    \hat{V}_0
    &= \frac{1}{N}\sum_{i=1}^N
        \{\hat{Y}_i(1)-\hat{Y}_i(0)-\hat{\tau}_0\}^2\\
        &\qquad + \frac{1}{N}\sum_{i=1}^N  \left[\left\{\frac{K_{M_0}(i)}{M_0}\right\}^2+\left(\frac{2M_0-1}{M_0}\right)\left\{\frac{K_{M_0}(i)}{M_0}\right\}\right] \hat{\sigma}^2(D_i, X_i) ,  \\
    \hat{V}_0^t
    &= \frac{1}{N_1}\sum_{\substack{1\le i\le N \\ D_i=1}}
        \{Y_i-\hat{Y}_i(0)-\hat{\tau}^t_0\}^2\\
        &\qquad+ \frac{1}{N_1}\sum_{i=1}^N
        (1-D_i)\left[\frac{K_{M_0}(i)\{K_{M_0}(i)-1\}}{M_0^2}\right]
        \hat{\sigma}^2(D_i, X_i) ,
\end{align*}
where $\hat{\sigma}^2(D_i, X_i)$ is defined in the main article.

\subsection{Martingale representation}

\cite{abadie2012martingale} provide a martingale representation for matching estimators.
This martingale representation holds no matter how matching is done.
According to \cite{abadie2012martingale}, this martingale representation is useful for analysing the asymptotic distribution of the estimator and correcting the standard error of a sample mean when missing data are imputed using the ``hot deck".
However, since they are beyond the scope of this article, we leave them for future study.

\clearpage
\bibliographystyle{chicago}
\bibliography{references}

@article{ivaRonaldSeema2006,
	author = {Iva Kozakova and Ronald Meester and Seema Nanda},
	journal = {The Annals of Probability},
	pages = {528--538},
	title = {The Size of Components in Continuum Nearest-Neighbor Graphs},
	volume = {34},
	year = {2006}}

@article{zong1998,
	author = {Zong, C.},
	journal = {Bulletin of the London Mathematical Society},
	pages = {1--10},
	title = {The kissing numbers of convex bodies---a brief survey},
	volume = {30},
	year = {1998}}

@article{CohnZhao2014,
	author = {Cohn, H and Zhao, Y.},
	journal = {Duke Math. J.},
	pages = {1965--2002},
	title = {Sphere packing bounds via spherical codes},
	volume = {163},
	year = {2014}}

@article{abadie2006large,
	author = {Abadie, Alberto and Imbens, Guido W},
	journal = {Econometrica},
	number = {1},
	pages = {235--267},
	publisher = {Wiley Online Library},
	title = {Large sample properties of matching estimators for average treatment effects},
	volume = {74},
	year = {2006}}

@article{abadie2011bias,
	author = {Abadie, Alberto and Imbens, Guido W},
	journal = {Journal of Business \& Economic Statistics},
	number = {1},
	pages = {1--11},
	publisher = {Taylor \& Francis},
	title = {Bias-corrected matching estimators for average treatment effects},
	volume = {29},
	year = {2011}}

@article{otsu2017bootstrap,
	author = {Otsu, Taisuke and Rai, Yoshiyasu},
	journal = {Journal of the American Statistical Association},
	number = {520},
	pages = {1720--1732},
	publisher = {Taylor \& Francis},
	title = {Bootstrap inference of matching estimators for average treatment effects},
	volume = {112},
	year = {2017}}

@article{abadie2008failure,
	author = {Abadie, Alberto and Imbens, Guido W},
	journal = {Econometrica},
	number = {6},
	pages = {1537--1557},
	publisher = {Wiley Online Library},
	title = {On the failure of the bootstrap for matching estimators},
	volume = {76},
	year = {2008}}

@article{rosenbaum2020modern,
	author = {Rosenbaum, Paul R},
	journal = {Annual Review of Statistics and Its Application},
	pages = {143--176},
	publisher = {Annual Reviews},
	title = {Modern algorithms for matching in observational studies},
	volume = {7},
	year = {2020}}

@article{imbens2009recent,
	author = {Imbens, Guido W and Wooldridge, Jeffrey M},
	journal = {Journal of Economic Literature},
	number = {1},
	pages = {5--86},
	title = {Recent developments in the econometrics of program evaluation},
	volume = {47},
	year = {2009}}

@article{rubin1974estimating,
	author = {Rubin, Donald B},
	journal = {Journal of Educational Psychology},
	number = {5},
	pages = {688-701},
	publisher = {American Psychological Association},
	title = {Estimating causal effects of treatments in randomized and nonrandomized studies.},
	volume = {66},
	year = {1974}}

@article{frolich2004finite,
	author = {Fr{\"o}lich, Markus},
	journal = {Review of Economics and Statistics},
	number = {1},
	pages = {77--90},
	publisher = {MIT Press 238 Main St., Suite 500, Cambridge, MA 02142-1046, USA journals~{\ldots}},
	title = {Finite-sample properties of propensity-score matching and weighting estimators},
	volume = {86},
	year = {2004}}

@article{busso2014new,
	author = {Busso, Matias and DiNardo, John and McCrary, Justin},
	journal = {Review of Economics and Statistics},
	number = {5},
	pages = {885--897},
	publisher = {The MIT Press},
	title = {New evidence on the finite sample properties of propensity score reweighting and matching estimators},
	volume = {96},
	year = {2014}}

@article{hahn1998role,
	author = {Hahn, Jinyong},
	journal = {Econometrica},
	pages = {315--331},
	publisher = {JSTOR},
	title = {On the role of the propensity score in efficient semiparametric estimation of average treatment effects},
	volume = {66},
	year = {1998}}

@article{rosenbaum1983central,
	author = {Rosenbaum, Paul R and Rubin, Donald B},
	journal = {Biometrika},
	number = {1},
	pages = {41--55},
	publisher = {Oxford University Press},
	title = {The central role of the propensity score in observational studies for causal effects},
	volume = {70},
	year = {1983}}

@unpublished{abadie2002simple,
	author = {Abadie, Alberto and Imbens, Guido},
	note = {manuscript},
	title = {Simple and bias-corrected matching estimators for average treatment effects},
	year = {2002}}

@article{abadie2012martingale,
	author = {Abadie, Alberto and Imbens, Guido W},
	journal = {Journal of the American Statistical Association},
	number = {498},
	pages = {833--843},
	publisher = {Taylor \& Francis},
	title = {A martingale representation for matching estimators},
	volume = {107},
	year = {2012}}

@article{lalonde1986evaluating,
	author = {LaLonde, Robert J},
	journal = {American Economic Review},
	pages = {604--620},
	publisher = {JSTOR},
	title = {Evaluating the econometric evaluations of training programs with experimental data},
	volume = {76},
	year = {1986}}

@article{dehejia1999causal,
	author = {Dehejia, Rajeev H and Wahba, Sadek},
	journal = {Journal of the American Statistical Association},
	number = {448},
	pages = {1053--1062},
	publisher = {Taylor \& Francis},
	title = {Causal effects in nonexperimental studies: Reevaluating the evaluation of training programs},
	volume = {94},
	year = {1999}}

@article{abadie2004implementing,
	author = {Abadie, Alberto and Drukker, David and Herr, Jane Leber and Imbens, Guido W},
	journal = {The Stata Journal},
	number = {3},
	pages = {290--311},
	publisher = {SAGE Publications Sage CA: Los Angeles, CA},
	title = {Implementing matching estimators for average treatment effects in Stata},
	volume = {4},
	year = {2004}}

@article{heckman1998matching,
	author = {Heckman, James J and Ichimura, Hidehiko and Todd, Petra},
	journal = {Review of Economic Studies},
	number = {2},
	pages = {261--294},
	publisher = {Wiley-Blackwell},
	title = {Matching as an econometric evaluation estimator},
	volume = {65},
	year = {1998}}

@article{hirano2003efficient,
	author = {Hirano, Keisuke and Imbens, Guido W and Ridder, Geert},
	journal = {Econometrica},
	number = {4},
	pages = {1161--1189},
	publisher = {Wiley Online Library},
	title = {Efficient estimation of average treatment effects using the estimated propensity score},
	volume = {71},
	year = {2003}}

@article{heckman1989choosing,
	author = {Heckman, James J and Hotz, V Joseph},
	journal = {Journal of the American Statistical Association},
	number = {408},
	pages = {862--874},
	publisher = {Taylor \& Francis},
	title = {Choosing among alternative nonexperimental methods for estimating the impact of social programs: The case of manpower training},
	volume = {84},
	year = {1989}}

@article{imbens2003sensitivity,
	author = {Imbens, Guido W},
	journal = {American Economic Review},
	number = {2},
	pages = {126--132},
	title = {Sensitivity to exogeneity assumptions in program evaluation},
	volume = {93},
	year = {2003}}

@article{smith2005does,
	author = {Smith, Jeffrey A and Todd, Petra E},
	journal = {Journal of Econometrics},
	number = {1-2},
	pages = {305--353},
	publisher = {Elsevier},
	title = {Does matching overcome {L}a{L}onde's critique of nonexperimental estimators?},
	volume = {125},
	year = {2005}}

@unpublished{imbens2005mean,
	author = {Imbens, Guido W and Newey, Whitney K and Ridder, Geert},
	note = {manuscript},
	title = {Mean-square-error calculations for average treatment effects},
	year = {2005}}

@article{chen2008semiparametric,
	author = {Chen, Xiaohong and Hong, Han and Tarozzi, Alessandro},
	journal = {Annals of Statistics},
	number = {2},
	pages = {808--843},
	publisher = {Institute of Mathematical Statistics},
	title = {Semiparametric efficiency in {GMM} models with auxiliary data},
	volume = {36},
	year = {2008}}

@article{horvitz1952generalization,
	author = {Horvitz, Daniel G and Thompson, Donovan J},
	journal = {Journal of the American Statistical Association},
	number = {260},
	pages = {663--685},
	publisher = {Taylor \& Francis},
	title = {A generalization of sampling without replacement from a finite universe},
	volume = {47},
	year = {1952}}

@article{robins1995semiparametric,
	author = {Robins, James M and Rotnitzky, Andrea},
	journal = {Journal of the American Statistical Association},
	number = {429},
	pages = {122--129},
	publisher = {Taylor \& Francis},
	title = {Semiparametric efficiency in multivariate regression models with missing data},
	volume = {90},
	year = {1995}}

@article{abadie2005semiparametric,
	author = {Abadie, Alberto},
	journal = {Review of Economic Studies},
	number = {1},
	pages = {1--19},
	publisher = {Wiley-Blackwell},
	title = {Semiparametric difference-in-differences estimators},
	volume = {72},
	year = {2005}}

@article{lin2023estimation,
	author = {Lin, Zhexiao and Ding, Peng and Han, Fang},
	journal = {Econometrica},
	number = {6},
	pages = {2187--2217},
	publisher = {Wiley Online Library},
	title = {Estimation based on nearest neighbor matching: from density ratio to average treatment effect},
	volume = {91},
	year = {2023}}

@article{diamond2013genetic,
	author = {Diamond, Alexis and Sekhon, Jasjeet S},
	journal = {Review of Economics and Statistics},
	number = {3},
	pages = {932--945},
	publisher = {The MIT Press},
	title = {Genetic matching for estimating causal effects: A general multivariate matching method for achieving balance in observational studies},
	volume = {95},
	year = {2013}}

@article{rubin1973matching,
	author = {Rubin, Donald B},
	journal = {Biometrics},
	pages = {159--183},
	publisher = {JSTOR},
	title = {Matching to remove bias in observational studies},
	volume = {29},
	year = {1973}}

@article{rubin1977assignment,
	author = {Rubin, Donald B},
	journal = {Journal of Educational Statistics},
	number = {1},
	pages = {1--26},
	publisher = {Sage Publications Sage CA: Thousand Oaks, CA},
	title = {Assignment to treatment group on the basis of a covariate},
	volume = {2},
	year = {1977}}

@article{rubin1973use,
	author = {Rubin, Donald B},
	journal = {Biometrics},
	pages = {185--203},
	publisher = {JSTOR},
	title = {The use of matched sampling and regression adjustment to remove bias in observational studies},
	volume = {29},
	year = {1973}}

@article{rubin1978bayesian,
	author = {Rubin, Donald B},
	journal = {Annals of Statistics},
	number = {1},
	pages = {34--58},
	publisher = {JSTOR},
	title = {Bayesian inference for causal effects: The role of randomization},
	volume = {6},
	year = {1978}}

@article{quade1982nonparametric,
	author = {Quade, Dana},
	journal = {Biometrics},
	pages = {597--611},
	publisher = {JSTOR},
	title = {Nonparametric analysis of covariance by matching},
	volume = {38},
	year = {1982}}
\end{document}